\documentclass[11pt,english]{article}
\usepackage[latin9]{inputenc}
\usepackage{amsmath}
\usepackage{amssymb}
\usepackage{graphicx}
\usepackage{setspace}
\makeatletter
\numberwithin{equation}{section}

\@ifundefined{date}{}{\date{}}
\usepackage{esint}
\usepackage{latexsym}
\usepackage{graphicx}\usepackage{bm}\usepackage{longtable}
\usepackage{hyperref}

\usepackage{xcolor}

\@addtoreset{equation}{section}

\makeatother

\usepackage{babel}
\begin{document}
\title{Out-of-time-order correlators of Skyrmion as baryon in holographic
QCD}
\maketitle
\begin{center}
Si-wen Li\footnote{Email: siwenli@dlmu.edu.cn}, Yi-peng Zhang\footnote{Email: ypmahler111@dlmu.edu.cn},
Hao-qian Li\footnote{Email: lihaoqian@dlmu.edu.cn},
\par\end{center}

\begin{center}
\emph{Department of Physics, School of Science,}\\
\emph{Dalian Maritime University, }\\
\emph{Dalian 116026, China}\\
\par\end{center}

\vspace{12mm}

\begin{abstract}
As the out-of-time-order correlator (OTOC) is a measure of quantum
chaos and an important observable in the context of AdS/CFT, we investigate
the OTOC of holographic Skyrmion which is described by an analytical
quantum mechanical system from the D4/D8 model (as the holographic
QCD). By employing the OTOC defined in quantum mechanics, we derive
the formulas and demonstrate the numerical calculations of the OTOC
explicitly which is also available for the general case with central
force field. Our numerical evaluation illustrates the behaviors of
the OTOC with large $N_{c}$, however the expected exponential growth
of OTOC is not obtained. Besides, we also take a look at the classical
limit of the OTOC and analyze the associated behaviors. At the end
of this work, we additionally study the OTOC with three-dimensional
Coulomb potential, as another example for the central force field,
to support our analyses of the general properties of quantum OTOC.
\end{abstract}
\newpage{}

\tableofcontents{}

\section{Introduction}

The out-of-time-order correlator (OTOC) $C_{T}\left(t\right)$ was
first introduced to calculate the vertex correction of a current for
a superconductor \cite{key-1}, which in general is defined as,

\begin{equation}
C_{T}=-\left\langle \left[W\left(t\right),V\left(0\right)\right]^{2}\right\rangle ,
\end{equation}
where $\left\langle ...\right\rangle $ refers to the thermal average.
And $W\left(t\right),V\left(t\right)$ are operators in the Heisenberg
picture at time $t$. In recent years, it trends to consider the OTOC
as a measure of the magnitude of quantum chaos \cite{key-2}. In particular,
for a quantum mechanical system, the OTOC is suggested to be defined
as \cite{key-3,key-4,key-5},

\begin{equation}
C_{T}=-\left\langle \left[x\left(t\right),p\left(0\right)\right]^{2}\right\rangle ,\label{eq:1.2}
\end{equation}
where $x,p$ refers to the canonical coordinate and momentum. In this
sense, by taking the semiclassical limit, i.e. replace the quantum
commutator $\frac{1}{i}\left[,\right]$ by the Poisson bracket $\left\{ ,\right\} _{\mathrm{P.B}}$,
the relation of classical OTOC and chaos can be found as

\begin{equation}
C_{T}\sim\left\{ x\left(t\right),p\left(0\right)\right\} _{\mathrm{P.B}}^{2}\sim\left[\frac{\delta x\left(t\right)}{\delta x\left(0\right)}\right]^{2}\sim e^{2Lt},\label{eq:1.3}
\end{equation}
for a classically chaotic system with the Lyapunov exponent $L\geq0$,
since such a system would be sensitive to the initial conditions.
However, the quantum version (\ref{eq:1.2}) of the OTOC defined in
(\ref{eq:1.3}) may not grow eternally, instead it could saturate
at the Ehrenfest time $t_{E}$. The Ehrenfest time $t_{E}$ refers
to the time scale beyond which the wave function spreads over the
whole system, thus it characterizes roughly the boundary between a
particle-like and a wave-like behavior of a wave function. Therefore,
it is conjectured that the exponent growth of the OTOC is only the
characteristic feature of the classical system.

Besides, the OTOC has also become to be an important observable in
the context of AdS/CFT correspondence, gauge-gravity duality \cite{key-6,key-7}
or quantum gravity. The upper bound of the quantum Lyapunov exponent
is proposed as $L\leq2\pi T$ \cite{key-2} which is originally discussed
in the context of quantum information on the horizon of black hole
\cite{key-8,key-9,key-10,key-11,key-12,key-13}. In the Sachdev-Ye-Kitaev
(SYK) model \cite{key-14,key-15} which is a quantum mechanical system
with infinitely long range disorder interactions of Majorana fermions,
the bound of quantum Lyapunov exponent is saturated \cite{key-16,key-17,key-18}.
Thus it strongly implies the SYK model can describe a quantum black
hole in holography or through AdS/CFT correspondence. Although the
presented works reveal several connections of OTOC and chaos in e.g.
quantum and black hole system, it is less clear whether the typical
chaotic systems display exponent growth of OTOC and whether the behavior
of OTOC is always consistent by taking into account the framework
of AdS/CFT or gauge-gravity duality.

Motivated by these problems, we would like to further investigate
the OTOC defined in (\ref{eq:1.2}) in quantum mechanics and moreover
in holography. Fortunately, the holographic quantum system describing
the Skyrmion in the D4/D8 model \cite{key-19,key-20,key-21}, as an
holographic baryonic system and as another holographic example distinct
from the SYK model, could be a good choice and sample to test the
quantum OTOC for our goal due to the following three reasons: 1) It
is a quantum mechanical system so that the OTOC defined in (\ref{eq:1.2})
should be valid. 2) It is a holographic system from the gauge-gravity
duality, thus it is possible to find the large N behavior of the OTOC
in this system. 3) The eigenvalue and eigenfunction of this system
are all analytical. 

Basically, the D4/D8 model (also named as Witten-Sakai-Sugimoto model)
is a very famous holographic model of QCD based on type IIA string
theory \cite{key-22,key-23,key-24}, since this model can almost include
all the elementary features of QCD by using the framework of string
theory, e.g. quark, meson \cite{key-19}, baryon \cite{key-20,key-21,key-25,key-26},
phase diagram \cite{key-27,key-28,key-29,key-30,key-31}, theta-dependence
\cite{key-32,key-33,key-34}, glueball \cite{key-35,key-36,key-37,key-38}.
In this work, we will focus on the baryon sector of the D4/D8 model
in which baryon is identified to a D4-brane wrapped on $S^{4}$ in
this model \cite{key-39}, however in the view of the D8-branes, such
a D4-brane is equivalent to the Yang-Mills instanton on the worldvolume
\cite{key-19,key-20,key-21,key-25,key-26,key-40}. Thus employing
the idea of Skyrmion \cite{key-41}, the collective excitation of
the instantonic soliton on the worldvolume of the D8-branes should
be recognized to the baryon state. On the other hand, this baryonic
system from the D4/D8 model must be strongly coupled (i.e. the t'
Hooft coupling constant $\lambda\rightarrow\infty$) according to
AdS/CFT dictionary, and in this limit, it turns out the Yang-Mills
instanton can be totally analytically solved by the Belavin--Polyakov--Schwarz--Tyupkin
(BPST) instanton solution, and its quantized version can also have
analytical eigenvalues and eigenfunctions which can been detailedly
reviewed in \cite{key-22,key-23,key-24,key-25,key-26}. Therefore
it would be interesting to test the behavior of the OTOC defined in
(\ref{eq:1.2}) by using the quantum mechanical system for Skyrmion
in holography.

The outline of this work is as follows. In Section 2, we collect the
relevant part of the holographic Skyrmion as baryon from the framework
of the D4/D8 model, then derive the formulas of the quantum OTOC of
this system. In Section 3, we numerically evaluate the OTOC in the
case of two and three flavors, at various temperature and including
its large N behaviors. In Section 4, we briefly derive the classical
limit of the OTOC and compare it with its quantum version. Summary
and discussion are given in the last section as Section 5. Besides,
in the Appendix A, we summarize the derivation for properties of the
spherical harmonic function in arbitrarily dimension, which is very
useful to evaluate the OTOC with the central force field. Moreover,
to support the analysis in this work, we additionally study the quantum
OTOC (\ref{eq:1.2}) with three-dimensional (3d) Coulomb potential
(which is typical atomic system) in the Appendix B, to attempt to
collect the universal features of OTOC in quantum mechanics with central
force field. In the Appendix C, we finally investigate the truncation
dependence of the OTOC presented in this work.

\section{The holographic setup}

\subsection{The holographic Skyrmion as baryon from the D4/D8 model}

Let us collect the relevant part of the D4/D8 model for the Skyrmion
as baryon in holography. The complete review can be found in \cite{key-19,key-20,key-21,key-22,key-23,key-24,key-25}.

The D4/D8 model consists of a stack of $N_{c}$ D4-branes as color
branes \footnote{In this work, ``large N'' refers to ``large $N_{c}$'' since we
denote the color number as $N_{c}$.}, $N_{f}$ pairs of coincident D8- and anti D8-branes as flavor and
anti-flavor branes, hence the dual theory as the worldvolume theory
on the D4- and D8-branes contains the gauge symmetry of $U\left(N_{c}\right)$
as color symmetry and $U\left(N_{f}\right)$ as flavor symmetry. In
particular, as the D8- and anti D8-branes take the opposite charge,
so the flavor symmetry on the D8-branes in this model is usually denoted
as $U\left(N_{f}\right)_{R}\times U\left(N_{f}\right)_{L}$ which
is recognized as the chiral symmetry in QCD \cite{key-19,key-23,key-24}.
Since our concern is the Skyrmion as baryon in this model, let us
focus on the flavor branes in this model. Recall the 5d classical
effective action for the flavor D8-brane \cite{key-19,key-20,key-21,key-25},
in low-energy region, it is given as,

\begin{align}
S_{\mathrm{D8}} & =S_{\mathrm{YM}}\left[\mathcal{A}\right]+S_{\mathrm{CS}}\left[\mathcal{A}\right],\nonumber \\
S_{\mathrm{YM}}\left[\mathcal{A}\right] & =-\kappa\int d^{4}xdz\mathrm{Tr}\left[\frac{1}{2}h\left(z\right)\mathcal{F}_{\mu\nu}^{2}+k\left(z\right)\mathcal{F}_{\mu z}^{2}\right],\nonumber \\
S_{\mathrm{CS}}\left[\mathcal{A}\right] & =\frac{N_{c}}{24\pi^{2}}\int\omega_{5}^{U\left(N_{f}\right)}\left(\mathcal{A}\right),\nonumber \\
h\left(z\right) & =\left(1+z^{2}\right)^{-1/3},k\left(z\right)=1+z^{2},\label{eq:1}
\end{align}
where the indices $\mu,\nu$ run over $0,1...3$ denoting the spacetime
$\mathbb{R}^{4}$ and the coordinate $z$ refers to the 5th dimensionless
Cartesian coordinate as the holographic direction. Note that the above
formulas are expressed in the unit of energy scale $M_{KK}=1$ which
is the scale of supersymmetry breaking. And the parameter $\kappa$,
't Hooft coupling constant $\lambda$ are given as,

\begin{equation}
\kappa=a\lambda N_{c},\ a=\frac{1}{216\pi^{3}},\lambda=g_{\mathrm{YM}}^{2}N_{c}.
\end{equation}
$\mathcal{A}$ refers to the $U\left(N_{f}\right)$ Yang-Mills gauge
potential associated to the gauge field strength $\mathcal{F}$ as
$\mathcal{F}=d\mathcal{A}+i\mathcal{A}\wedge\mathcal{A}$ and the
gauge-invariant Chern-Simons (CS) 5-form $\omega_{5}^{U\left(N_{f}\right)}\left(\mathcal{A}\right)$
is given as,

\begin{equation}
\omega_{5}^{U\left(N_{f}\right)}\left(\mathcal{A}\right)=\mathrm{Tr}\left(\mathcal{A}\mathcal{F}^{2}-\frac{i}{2}\mathcal{A}^{3}\mathcal{F}-\frac{1}{10}\mathcal{A}^{5}\right).
\end{equation}
Since in the D4/D8 model the baryon vertex is identified as a D4-brane
wrapped on $S^{4}$ according to gauge-gravity duality \cite{key-39},
it is recognized to be equivalently as the instanton configuration
of the gauge field on the flavor D8-branes \cite{key-40}. Therefore,
it is possible to use the classical Yang-Mills instanton solution
as baryon in this model \cite{key-19,key-20,key-21} which, for generic
$N_{f}\geq2$, can be chosen as,

\begin{equation}
A_{M}^{\mathrm{cl}}=-if\left(\xi\right)g\left(x\right)\partial_{M}g^{-1},\label{eq:4}
\end{equation}
where

\begin{align}
f\left(\xi\right) & =\frac{\xi^{2}}{\xi^{2}+\rho^{2}},\xi=\sqrt{\left(x^{M}-X^{M}\right)^{2}},\nonumber \\
g\left(x\right) & =\left(\begin{array}{cc}
g^{SU\left(2\right)}\left(x\right) & 0\\
0 & \mathbf{1}_{N_{f}-2}
\end{array}\right),g^{SU\left(2\right)}\left(x\right)=\frac{1}{\xi}\left[\left(z-Z\right)\mathbf{1}_{2}-i\left(x^{i}-X^{i}\right)\tau^{i}\right],
\end{align}
and

\begin{align}
\hat{A}_{0}^{\mathrm{cl}} & =\sqrt{\frac{2}{N_{f}}}\frac{1}{8\pi^{2}a}\frac{1}{\xi^{2}}\left[1-\frac{\rho^{4}}{\left(\xi^{2}+\rho^{2}\right)^{2}}\right],\ \hat{A}_{M}^{\mathrm{cl}}=0,\nonumber \\
A_{0}^{\mathrm{cl}} & =\frac{1}{16\pi^{2}a}\frac{1}{\xi^{2}}\left[1-\frac{\rho^{4}}{\left(\xi^{2}+\rho^{2}\right)^{2}}\right]\left(\mathcal{P}_{2}-\frac{2}{N_{f}}\mathbf{1}_{N_{f}}\right).\label{eq:6}
\end{align}
Here the index $M,N$ runs over $1,2,3,z$ and $\mathcal{P}_{2}$
is an $N_{f}\times N_{f}$ matrix defined as $\mathcal{P}_{2}=\mathrm{diag}\left\{ 1,1,0,...0\right\} $.
Note that we have decomposed respectively the $U\left(N_{f}\right)$
group as $U\left(N_{f}\right)\simeq U\left(1\right)\times SU\left(N_{f}\right)$
and corresponding generator $\mathcal{A}$ as, 
\begin{equation}
\mathcal{A}=A+\frac{1}{\sqrt{2N_{f}}}\hat{A}=A^{a}t^{a}+\frac{1}{\sqrt{2N_{f}}}\hat{A},
\end{equation}
where $\hat{A},A$ refers respectively to the generator of $U\left(1\right)$,
$SU\left(N_{f}\right)$. The $N_{f}^{2}-1$ matrices $t^{a}$ ($a=1,2...N_{f}^{2}-1$)
are the normalized Hermitian bases of the $su\left(N_{f}\right)$
algebra satisfying

\begin{equation}
\mathrm{Tr}\left(t^{a}t^{b}\right)=\frac{1}{2}\delta^{ab}.
\end{equation}
The symbol $\mathbf{1}_{N}$ refers to the $N\times N$ identity matrix,
and $\tau^{i}$'s are the Pauli matrices. The constants $X^{M}=\left\{ X^{i},Z\right\} $
and $\rho$ refer respectively to the position and the size of the
instanton scaled in the large $\lambda$ limit. Following the discussion
in \cite{key-19,key-20,key-21}, one can confirm that the instanton
solution given in (\ref{eq:4}) - (\ref{eq:6}) satisfies the leading
order equations of motion obtained by varying action (\ref{eq:1})
in the large $\lambda$ limit, and the instanton solution (\ref{eq:4})
- (\ref{eq:6}) may also return to the classical Belavin--Polyakov--Schwarz--Tyupkin
(BPST) instanton solution for $N_{f}=2$.

Keeping the above in hand then employing the idea of Skyrmion \cite{key-41},
the baryon states are recognized to be the excitation of the collective
modes of the instantonic soliton described by the above BPST solution
\cite{key-19,key-20,key-21}. Thus the energy of the classical instantonic
soliton $M^{\mathrm{soliton}}$ can be evaluated by inserting the
classical instanton solution (\ref{eq:4}) - (\ref{eq:6}) back to
the action (\ref{eq:1}) with $S\left[\mathcal{A}^{\mathrm{cl}}\right]=-\int dtM^{\mathrm{soliton}}$.
Afterwards, in order to find the excitation of the collective modes,
let us slowly move the classical soliton, hence the collective coordinates
of the soliton are promoted to become time-dependent \cite{key-42}.
As the Lagrangian of the collective coordinates is expected to be
the element of the world line with a potential in the moduli space,
it is possible to obtain its quantized version by replacing momentum
to the associated derivative operator \cite{key-19,key-20,key-21}.
Altogether, we can reach to the following quantum mechanical system
to describe the soliton dynamics (the dynamics of Skyrmion) in the
flavored moduli space with the following Hamiltonian $H$,

\begin{align}
H= & M_{0}+H_{Z}+H_{y},\nonumber \\
H_{Z}= & -\frac{1}{2m_{Z}}\partial_{Z}^{2}+\frac{1}{2}m_{Z}\omega_{Z}^{2}Z^{2},\nonumber \\
H_{y}= & -\frac{1}{2m_{\rho}}\sum_{I=1}^{n+1}\frac{\partial^{2}}{\partial y_{I}^{2}}+\frac{1}{2}m_{\rho}\omega_{\rho}^{2}\rho^{2}+\frac{K}{m_{\rho}\rho^{2}},\label{eq:9}
\end{align}
where $\rho$ is the size of the instanton and $y_{I}$ refers to
the Cartesian coordinate in the moduli space satisfying

\begin{equation}
\sum_{I=1}^{n+1}y_{I}^{2}=\rho^{2}.\label{eq:10}
\end{equation}
In the unit of $M_{KK}=1$, the parameters are given as,
\begin{equation}
M_{0}=8\pi^{2}\kappa,m_{Z}=\frac{m_{\rho}}{2}=8\pi^{2}\kappa\lambda^{-1},K=\frac{2}{5}N_{c}^{2},\omega_{Z}^{2}=4\omega_{\rho}^{2}=\frac{2}{3}.
\end{equation}
Note that $n$ relates to the number of the generators of $SU\left(N_{f}\right)$
i.e. $n=N_{f}^{2}-1$. Therefore the collective modes of the soliton
are identified to the eigenstates of the Hamiltonian, i.e. the Skyrmion,
given in (\ref{eq:9}) which are the baryon states accordingly. Remarkably,
the Hamiltonian given in (\ref{eq:9}) is totally analytical. Using
the radial coordinate given in (\ref{eq:10}) , we find

\begin{equation}
H_{y}=-\frac{1}{2m_{\rho}}\left[\frac{1}{\rho^{n}}\partial_{\rho}\left(\rho^{n}\partial_{\rho}\right)+\frac{1}{\rho^{2}}\nabla_{S^{n}}^{2}\right]+\frac{1}{2}m_{\rho}\omega_{\rho}^{2}\rho^{2}+\frac{K}{m_{\rho}\rho^{2}},
\end{equation}
where $\nabla_{S^{n}}^{2}$ refers to the Laplacian operator on $S^{n}$.
Hence the eigenfunction of Hamiltonian (\ref{eq:9}) can be expressed
as,

\begin{equation}
\psi_{n_{Z},n_{\rho},l_{n},l_{n-1},...l_{1}}\left(Z,\rho,\theta_{n},\theta_{n-1},...\theta_{1}\right)=\chi_{n_{Z}}\left(Z\right)\mathcal{R}_{n_{\rho},l_{n}}\left(\rho\right)\mathcal{Y}\left(\theta_{n},\theta_{n-1},...\theta_{1}\right),
\end{equation}
where $\chi_{n_{Z}}\left(Z\right)$ refers to the eigenfunction of
$H_{Z}$ which is nothing but the eigenfunction of one-dimensional
harmonic oscillator. $\mathcal{Y}\left(\theta_{n},\theta_{n-1},...\theta_{1}\right)$
is the spherical harmonic function on $S^{n}$ satisfying (See the
details in the Appendix A),

\begin{equation}
\nabla_{S^{n}}^{2}\mathcal{Y}_{l_{n},l_{n-1},...l_{1}}\left(\theta_{n},\theta_{n-1},...\theta_{1}\right)=-l_{n}\left(l_{n}+n-1\right)\mathcal{Y}_{l_{n},l_{n-1},...l_{1}}\left(\theta_{n},\theta_{n-1},...\theta_{1}\right),
\end{equation}
$\mathcal{R}_{n_{\rho},l_{n}}\left(\rho\right)$ is given by the confluent
hypergeometric function $F\left(a,b,x\right)$ as,

\begin{align}
\mathcal{R}_{n_{\rho},l_{n}}\left(\rho\right) & =\mathcal{N}\left(n_{\rho},l_{n}\right)e^{-\frac{m_{\rho}\omega_{\rho}\rho^{2}}{2}}\rho^{\xi}F\left(-n_{\rho},\xi+\frac{n+1}{2},m_{\rho}\omega_{\rho}\rho^{2}\right),\nonumber \\
\xi & =\sqrt{2K+\left(l_{n}+\frac{n-1}{2}\right)^{2}}-\frac{n-1}{2},
\end{align}
with the normalization

\begin{equation}
\int d\rho\rho^{n}\mathcal{R}_{n_{\rho}^{\prime},l_{n}^{\prime}}\mathcal{R}_{n_{\rho},l_{n}}=\delta_{n_{\rho}^{\prime},n_{\rho}}.
\end{equation}
Afterwards, the eigenvalue of (\ref{eq:9}) is computed as,

\begin{equation}
E_{n_{Z},n_{\rho},l_{n}}\left(N_{c},n\right)=E_{n_{Z}}+E_{n_{\rho},l_{n}}\left(N_{c},n\right)+M_{0},\label{eq:17}
\end{equation}
where

\begin{equation}
E_{n_{Z}}=\left(n_{Z}+\frac{1}{2}\right)\omega_{Z},\ E_{n_{\rho},l_{n}}\left(N_{c},n\right)=\left[\sqrt{2K+\left(l_{n}+\frac{n-1}{2}\right)^{2}}+2n_{\rho}+1\right]\omega_{\rho}.
\end{equation}

\subsection{The general formulas of OTOC for the holographic Skyrmion}

In quantum mechanical system, for a time-independent Hamiltonian $H\left(x_{1},x_{2}...x_{J},p_{1},p_{2}...p_{J}\right)$,
\cite{key-3,key-4,key-5} suggest the thermal and microcanonical OTOC
$C_{T},c_{n}$ can be defined as,

\begin{equation}
C_{T}\left(t\right)=\frac{1}{\mathcal{Z}}\sum_{n}e^{-\beta E_{n}}c_{n}\left(t\right),\ \mathcal{Z}=\sum_{n}e^{-\beta E_{n}},\ c_{n}\left(t\right)=-\left\langle n\left|\left[x\left(t\right),p\right]^{2}\right|n\right\rangle ,\label{eq:19}
\end{equation}
where $x\left(t\right),p\left(t\right)$ is respectively the canonical
coordinate and momentum in Heisenberg picture, $\left|n\right\rangle $
refers to the $n$-th eigenstate of Hamiltonian satisfying $H\left|n\right\rangle =E_{n}\left|n\right\rangle $,
$\mathcal{Z}$ is the partition function. For notational simplicity,
we have denoted

\begin{equation}
x\left(t\right)=x_{a}\left(t\right),p\left(t\right)=p_{a}\left(t\right),p=p\left(0\right),a\in\left\{ 1,2...J\right\} \label{eq:20}
\end{equation}
Note that $\beta=1/T$ refers to the temperature of the system. For
the following numerical calculations, we can write $c_{n}\left(t\right)$
as,

\begin{equation}
c_{n}\left(t\right)=-\sum_{m}\left\langle n\left|\left[x\left(t\right),p\right]\right|m\right\rangle \left\langle m\left|\left[x\left(t\right),p\right]\right|n\right\rangle \equiv\sum_{m}b_{nm}\left(t\right)b_{nm}^{*}\left(t\right),\label{eq:21}
\end{equation}
where

\begin{equation}
b_{nm}\left(t\right)=-i\left\langle n\left|\left[x\left(t\right),p\right]\right|m\right\rangle .\label{eq:22}
\end{equation}
Recall the unitary transformation $x\left(t\right)=e^{iHt}xe^{-iHt}$,
(\ref{eq:22}) can be rewritten as,

\begin{equation}
b_{nm}\left(t\right)=-i\sum_{k}\left(e^{iE_{nk}t}x_{nk}p_{km}-e^{iE_{km}t}p_{nk}x_{km}\right),\label{eq:23}
\end{equation}
where $x_{nk}=\left\langle n\left|x\right|k\right\rangle ,p_{nk}=\left\langle n\left|p\right|k\right\rangle $
and $E_{nk}=E_{n}-E_{k}$. Notice for a time-independent Hamiltonian

\begin{equation}
H=\sum_{a=1}^{J}p_{a}^{2}+U\left(x_{1},x_{2}...x_{J}\right),
\end{equation}
we can obtain 
\begin{equation}
p_{mn}=\frac{i}{2}E_{mn}x_{mn},\label{eq:25}
\end{equation}
due to the commutation relation $\left[H,x\right]=-2ip$. Substituting
(\ref{eq:25}) into (\ref{eq:23}), it leads to

\begin{equation}
b_{nm}\left(t\right)=\frac{1}{2}\sum_{k}x_{nk}x_{km}\left(E_{km}e^{iE_{nk}t}-E_{nk}e^{iE_{km}t}\right).\label{eq:26}
\end{equation}
Therefore the OTOC can be computed by using (\ref{eq:19}) (\ref{eq:21})
and (\ref{eq:26}) once we obtain the matrix elements of $x$.

Next, let us apply the above formulas to our quantum mechanical system
presented in Section 2.1. Recall the Hamiltonian given in (\ref{eq:9})
which is defined in an $n+2$ dimensional space parameterized by the
collective coordinates $\left\{ y_{1},y_{2}...y_{n+1},Z\right\} $.
Notice that $H_{Z}$ is exactly the Hamiltonian of one-dimensional
harmonic oscillator, therefore in Heisenberg picture, the operator
$Z\left(t\right)$ and its associated momentum $p_{Z}\left(t\right)$
can be solved analytically from the Heisenberg equation as,

\begin{align}
Z\left(t\right) & =Z\cos\omega_{Z}t+\frac{p_{Z}}{m_{Z}}\sin\omega_{Z}t,\nonumber \\
p_{Z}\left(t\right) & =-m_{Z}\omega_{Z}\sin\omega_{Z}t+p_{Z}\omega_{Z}\cos\omega_{Z}t,\label{eq:27}
\end{align}
which means

\begin{equation}
\left[Z\left(t\right),p_{Z}\right]=\left[Z\cos\omega t,p_{Z}\right]=i\cos\omega_{Z}t,
\end{equation}
i.e. the OTOC $c_{n}^{\left(Z\right)}$for $Z$ part is

\begin{equation}
c_{n}^{\left(Z\right)}\left(t\right)=-\left\langle n\left|\left[Z\left(t\right),p_{Z}\right]^{2}\right|n\right\rangle =\cos^{2}\omega_{Z}t.
\end{equation}
Since the properties of the one-dimensional harmonic oscillator's
OTOC has been discussed in \cite{key-3,key-4}, we will focus on the
OTOC with respect to $H_{y}$ given in (\ref{eq:9}) instead of $H_{Z}$.
Denoting $x^{a}\in\left\{ y_{1},y_{2}...y_{n+1}\right\} $, we notice
that $H_{y}$ is spherically symmetric for $\left\{ y_{1},y_{2}...y_{n+1}\right\} $,
therefore we can chose arbitrary $x^{a}$ to compute the OTOC by following
(\ref{eq:19}) (\ref{eq:20}). In order to simplify the calculation,
we chose the $n+1$-th coordinate $x^{n+1}=y_{n+1}$ to compute its
matrix element. Using the spherical coordinate given in (\ref{eq:10}),
we have $x^{n+1}=\rho\cos\theta_{n}$ where $\theta_{n}$ refers to
the $n$-th polar angle of the $n+1$-dimensional spherical coordinate
(the details of the spherical coordinates in arbitrary dimension are
summarized in the Appendix A.) For the notational simplicity, we omit
the index $n+1$ of the $n+1$-th coordinate $x^{n+1}$ then its matrix
elements are given as,

\begin{align}
 & x_{n_{Z},n_{\rho},l_{n},l_{n-1},...l_{1};n_{Z}^{\prime},n_{\rho}^{\prime},l_{n}^{\prime},l_{n-1}^{\prime},...l_{1}^{\prime}}\nonumber \\
= & \left\langle n_{Z},n_{\rho},l_{n},l_{n-1},...l_{1}\left|\rho\cos\theta_{n}\right|n_{Z}^{\prime},n_{\rho}^{\prime},l_{n}^{\prime},l_{n-1}^{\prime},...l_{1}^{\prime}\right\rangle \nonumber \\
= & \delta_{n_{Z}^{\prime},n_{Z}}\rho_{n_{\rho}^{\prime},l_{n}^{\prime};n_{\rho},l_{n}}Y_{l_{n}^{\prime},l_{n-1}^{\prime},...l_{1}^{\prime};l_{n},l_{n-1},...l_{1}},
\end{align}
where

\begin{align}
\rho_{n_{\rho}^{\prime},l_{n}^{\prime};n_{\rho},l_{n}} & =\int\rho^{n+1}d\rho\mathcal{R}_{n_{\rho}^{\prime},l_{n}^{\prime}}\mathcal{R}_{n_{\rho},l_{n}},\nonumber \\
Y_{l_{n}^{\prime},l_{n-1}^{\prime},...l_{1}^{\prime};l_{n},l_{n-1},...l_{1}} & =\int dV_{S^{n}}\mathcal{Y}_{l_{n}^{\prime},l_{n-1}^{\prime},...l_{1}^{\prime}}^{*}\cos\theta_{n}\mathcal{Y}_{l_{n},l_{n-1},...l_{1}}.
\end{align}
Note that while the matrix element $\rho_{n_{\rho}^{\prime},l_{n}^{\prime};n_{\rho},l_{n}}$
must be evaluated numerically, the spherical matrix element $Y_{l_{n}^{\prime},l_{n-1}^{\prime},...l_{1}^{\prime};l_{n},l_{n-1},...l_{1}}$
is analytically as it is given in the Appendix A. To simplify the
calculations, let us define the kernel of the functions as,

\begin{align}
Y_{l_{n}^{\prime},l_{n-1}^{\prime},...l_{1}^{\prime};l_{n},l_{n-1},...l_{1}} & =Y_{l_{n}^{\prime},l_{n-1}^{\prime};l_{n},l_{n-1}}^{\mathrm{ker}}\delta_{n_{Z}^{\prime},n_{Z}}\prod_{j=1}^{n-2}\delta_{l_{j}^{\prime},l_{j}},\nonumber \\
x_{n_{\rho}^{\prime},l_{n}^{\prime},l_{n-1}^{\prime};n_{\rho},l_{n},l_{n-1}}^{\mathrm{ker}} & =\rho_{n_{\rho}^{\prime},l_{n}^{\prime};n_{\rho},l_{n}}Y_{l_{n}^{\prime},l_{n-1}^{\prime};l_{n},l_{n-1}}^{\mathrm{ker}}.
\end{align}
where according to the Appendix A, the kernel of the functions are
computed as,

\begin{align}
Y_{l_{n}^{\prime},l_{n-1}^{\prime};l_{n},l_{n-1}}^{\mathrm{ker}}= & \bigg[\sqrt{\frac{\left(l_{n}-l_{n-1}\right)\left(l_{n}+l_{n-1}+n-2\right)}{\left(2l_{n}+n-1\right)\left(2l_{n}+n-3\right)}}\delta_{l_{n}^{\prime},l_{n}-1}\nonumber \\
 & +\sqrt{\frac{\left(l_{n}+l_{n-1}+n-1\right)}{\left(2l_{n}+n-1\right)}\frac{\left(l_{n}-l_{n-1}+1\right)}{\left(2l_{n}+n+1\right)}}\delta_{l_{n}^{\prime},l_{n}+1}\bigg]\delta_{l_{n-1}^{\prime},l_{n-1}}.\label{eq:2.33}
\end{align}
Thus we get the matrix elements of $b_{nm}$ defined in (\ref{eq:26})
as,

\begin{align}
 & b_{n_{Z},n_{\rho},l_{n},l_{n-1},...l_{1};n_{Z}^{\prime},n_{\rho}^{\prime},l_{n}^{\prime},l_{n-1}^{\prime},...l_{1}^{\prime}}\left(t\right)\nonumber \\
= & \frac{1}{2}\sum_{n_{\rho}^{\prime\prime},l_{n}^{\prime\prime},l_{n-1}^{\prime\prime}}\delta_{n_{Z},n_{Z}^{\prime}}x_{n_{\rho},l_{n},l_{n-1};n_{\rho}^{\prime\prime},l_{n}^{\prime\prime},l_{n-1}^{\prime\prime}}^{\mathrm{ker}}x_{n_{\rho}^{\prime\prime},l_{n}^{\prime\prime},l_{n-1}^{\prime\prime};n_{\rho}^{\prime},l_{n}^{\prime},l_{n-1}^{\prime}}^{\mathrm{ker}}\prod_{j=1}^{n-2}\delta_{l_{j},l_{j}^{\prime}}\nonumber \\
 & \times\bigg(E_{n_{Z}^{\prime},n_{\rho}^{\prime\prime},l_{n}^{\prime\prime};n_{Z}^{\prime},n_{\rho}^{\prime},l_{n}^{\prime}}e^{iE_{n_{Z},n_{\rho},l_{n};n_{Z}^{\prime},n_{\rho}^{\prime\prime},l_{n}^{\prime\prime}}t}-E_{n_{Z},n_{\rho},l_{n};n_{Z}^{\prime},n_{\rho}^{\prime\prime},l_{n}^{\prime\prime}}e^{iE_{n_{Z}^{\prime},n_{\rho}^{\prime\prime},l_{n}^{\prime\prime};n_{Z}^{\prime},n_{\rho}^{\prime},l_{n}^{\prime}}t}\bigg).
\end{align}
so that the microcanonical OTOC $c_{n}\left(t\right)$ can be written
as,

\begin{align}
 & c_{n_{Z},n_{\rho},l_{n},l_{n-1},...l_{1}}\left(t\right)\nonumber \\
= & \sum_{n_{Z}^{\prime},n_{\rho}^{\prime},l_{n}^{\prime},l_{n-1}^{\prime},...l_{1}^{\prime}}b_{n_{Z},n_{\rho},l_{n},l_{n-1},...l_{1};n_{Z}^{\prime},n_{\rho}^{\prime},l_{n}^{\prime},l_{n-1}^{\prime},...l_{1}^{\prime}}^{*}b_{n_{Z},n_{\rho},l_{n},l_{n-1},...l_{1};n_{Z}^{\prime},n_{\rho}^{\prime},l_{n}^{\prime},l_{n-1}^{\prime},...l_{1}^{\prime}}\nonumber \\
= & \frac{1}{4}\sum_{n_{\rho}^{\prime},l_{n}^{\prime},l_{n-1}^{\prime};n_{\rho}^{\prime\prime},l_{n}^{\prime\prime},l_{n-1}^{\prime\prime};n_{\rho}^{\prime\prime\prime},l_{n}^{\prime\prime\prime},l_{n-1}^{\prime\prime\prime}}\nonumber \\
 & \bigg[x_{n_{\rho},l_{n},l_{n-1};n_{\rho}^{\prime\prime},l_{n}^{\prime\prime},l_{n-1}^{\prime\prime}}^{\mathrm{ker}}x_{n_{\rho}^{\prime\prime},l_{n}^{\prime\prime},l_{n-1}^{\prime\prime};n_{\rho}^{\prime},l_{n}^{\prime},l_{n-1}^{\prime}}^{\mathrm{ker}}x_{n_{\rho},l_{n},l_{n-1};n_{\rho}^{\prime\prime\prime},l_{n}^{\prime\prime\prime},l_{n-1}^{\prime\prime\prime}}^{\mathrm{ker}}x_{n_{\rho}^{\prime\prime\prime},l_{n}^{\prime\prime\prime},l_{n-1}^{\prime\prime\prime};n_{\rho}^{\prime},l_{n}^{\prime},l_{n-1}^{\prime}}^{\mathrm{ker}}\nonumber \\
 & \times\bigg(E_{n_{\rho}^{\prime\prime},l_{n}^{\prime\prime};n_{\rho}^{\prime},l_{n}^{\prime}}e^{-iE_{n_{\rho},l_{n};n_{\rho}^{\prime\prime},l_{n}^{\prime\prime}}t}-E_{n_{\rho},l_{n};n_{\rho}^{\prime\prime},l_{n}^{\prime\prime}}e^{-iE_{n_{\rho}^{\prime\prime},l_{n}^{\prime\prime};n_{\rho}^{\prime},l_{n}^{\prime}}t}\bigg)\nonumber \\
 & \times\bigg(E_{n_{\rho}^{\prime\prime\prime},l_{n}^{\prime\prime\prime};n_{\rho}^{\prime},l_{n}^{\prime}}e^{iE_{n_{\rho},l_{n};n_{\rho}^{\prime\prime\prime},l_{n}^{\prime\prime\prime}}t}-E_{n_{\rho},l_{n};n_{\rho}^{\prime\prime\prime},l_{n}^{\prime\prime\prime}}e^{iE_{n_{\rho}^{\prime\prime\prime},l_{n}^{\prime\prime\prime};n_{\rho}^{\prime},l_{n}^{\prime}}t}\bigg)\bigg],\label{eq:35}
\end{align}
where
\begin{equation}
E_{n_{Z},n_{\rho},l_{n};n_{Z},n_{\rho}^{\prime},l_{n}^{\prime}}\equiv E_{n_{\rho},l_{n}}-E_{n_{\rho}^{\prime},l_{n}^{\prime}}=E_{n_{\rho},l_{n};n_{\rho}^{\prime},l_{n}^{\prime}}.
\end{equation}
Altogether, the OTOC would be evaluated numerically once we obtain
the kernel of the matrix elements $x_{n_{\rho}^{\prime\prime},l_{n}^{\prime\prime},l_{n-1}^{\prime\prime};n_{\rho}^{\prime},l_{n}^{\prime},l_{n-1}^{\prime}}^{\mathrm{ker}}$.
Besides, we can find the microcanonical OTOC depends on the quantum
numbers $n_{\rho},l_{n},l_{n-1}$ only. So the thermal OTOC can be
obtained by,

\begin{equation}
C_{T}\left(t\right)=\frac{1}{\mathcal{Z}}\sum_{n_{Z},n_{\rho},l_{n},l_{n-1}...l_{1}}c_{n_{Z},n_{\rho},l_{n},l_{n-1},...l_{1}}\left(t\right)e^{-\frac{E_{n_{Z},n_{\rho},l_{n}}}{T}},\label{eq:37}
\end{equation}
where the partition function is given by

\begin{equation}
\mathcal{Z}=\sum_{n_{Z},n_{\rho},l_{n},l_{n-1}...l_{1}}e^{-\frac{E_{n_{Z},n_{\rho},l_{n}}}{T}}=\sum_{n_{Z},n_{\rho},l_{n}}n_{\mathrm{d}}e^{-\frac{E_{n_{Z},n_{\rho},l_{n}}}{T}},\label{eq:38}
\end{equation}
and 
\begin{equation}
n_{\mathrm{d}}\left(l_{n}\right)=\sum_{l_{n-1}=0}^{l_{n}}\sum_{l_{n-2}=0}^{l_{n-1}}...\sum_{l_{3}=0}^{l_{4}}\sum_{l_{2}=0}^{l_{3}}\left(2l_{2}+1\right),
\end{equation}
is the degeneracy number for a given energy $E_{n_{Z},n_{\rho},l_{n}}$.
We will numerically evaluate the quantum OTOC by using (\ref{eq:35})
- (\ref{eq:38}) in the next section.

\section{The numerical analysis}

\subsection{Case of two flavors}

\begin{figure}[t]
\begin{centering}
\includegraphics[scale=0.37]{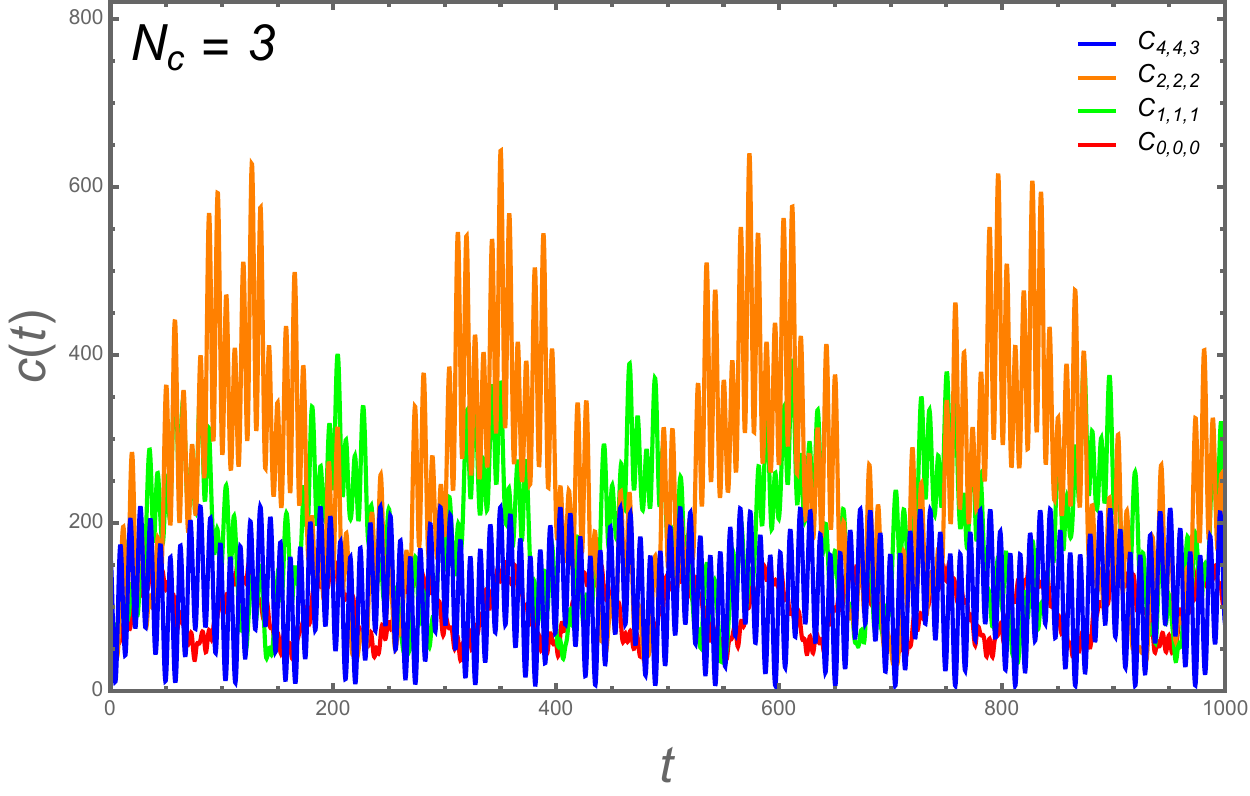}\includegraphics[scale=0.37]{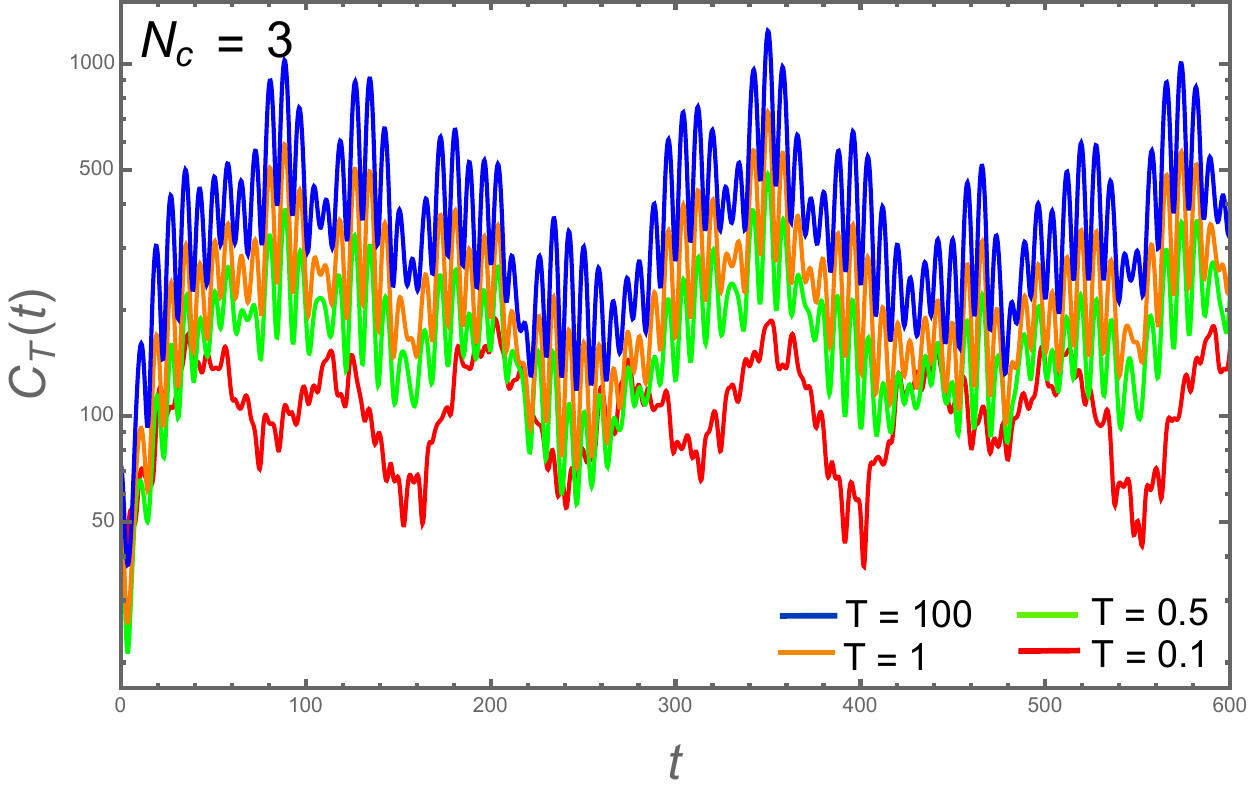}
\par\end{centering}
\begin{centering}
\includegraphics[scale=0.37]{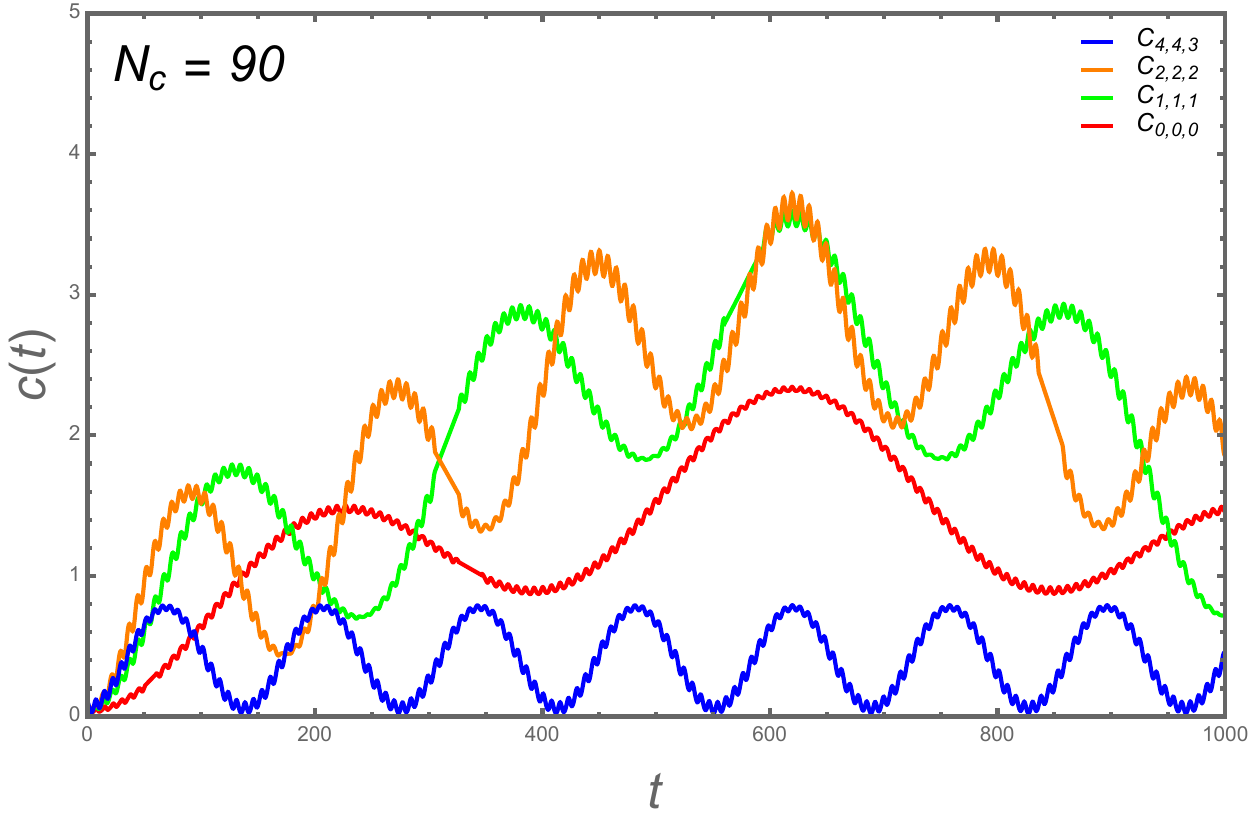}\includegraphics[scale=0.37]{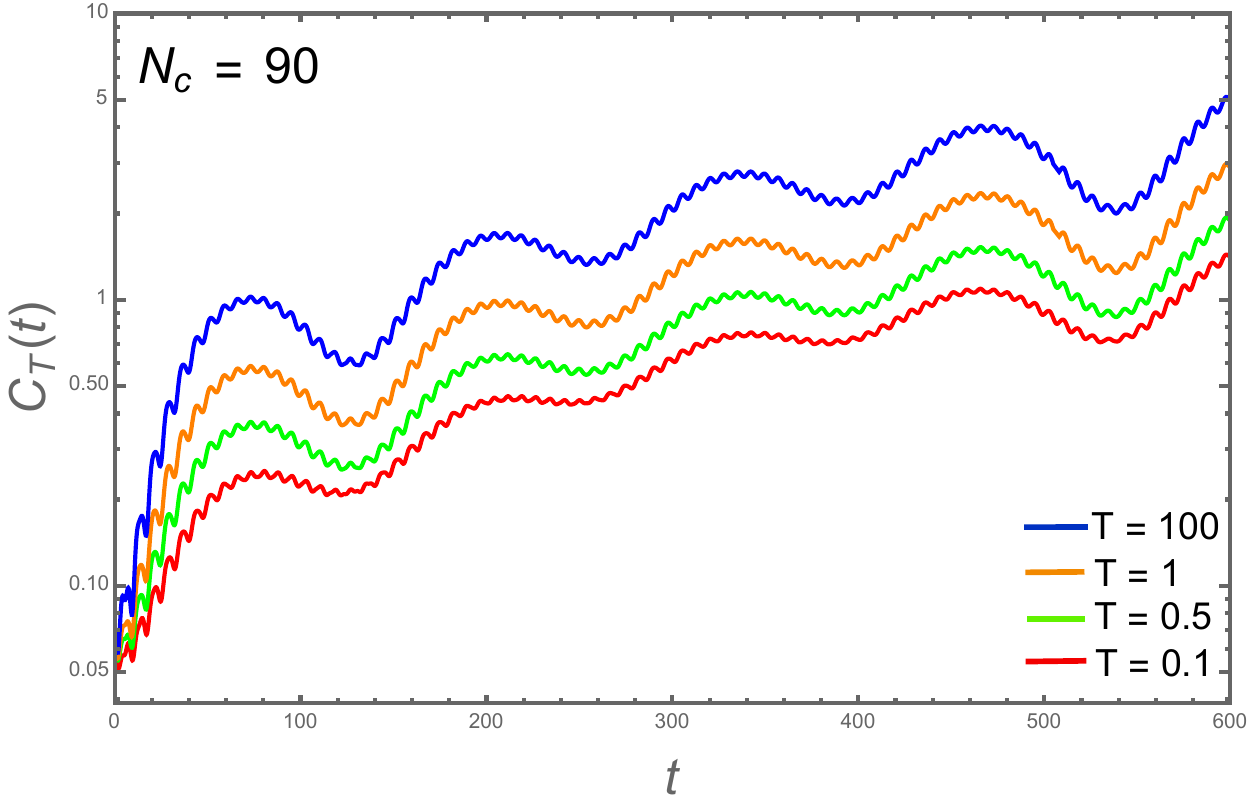}
\par\end{centering}
\caption{\label{fig:1} Microcanonical and thermal OTOC of holographic Skyrmion
with two flavors $N_{f}=2$. \textbf{Left:} Microcanonical OTOC $c\left(t\right)$
as a function of time $t$. \textbf{Right:} Thermal OTOC $C_{T}\left(t\right)$
as a function of time $t$ and temperature $T$.}
\end{figure}
\begin{figure}[t]
\begin{centering}
\includegraphics[scale=0.37]{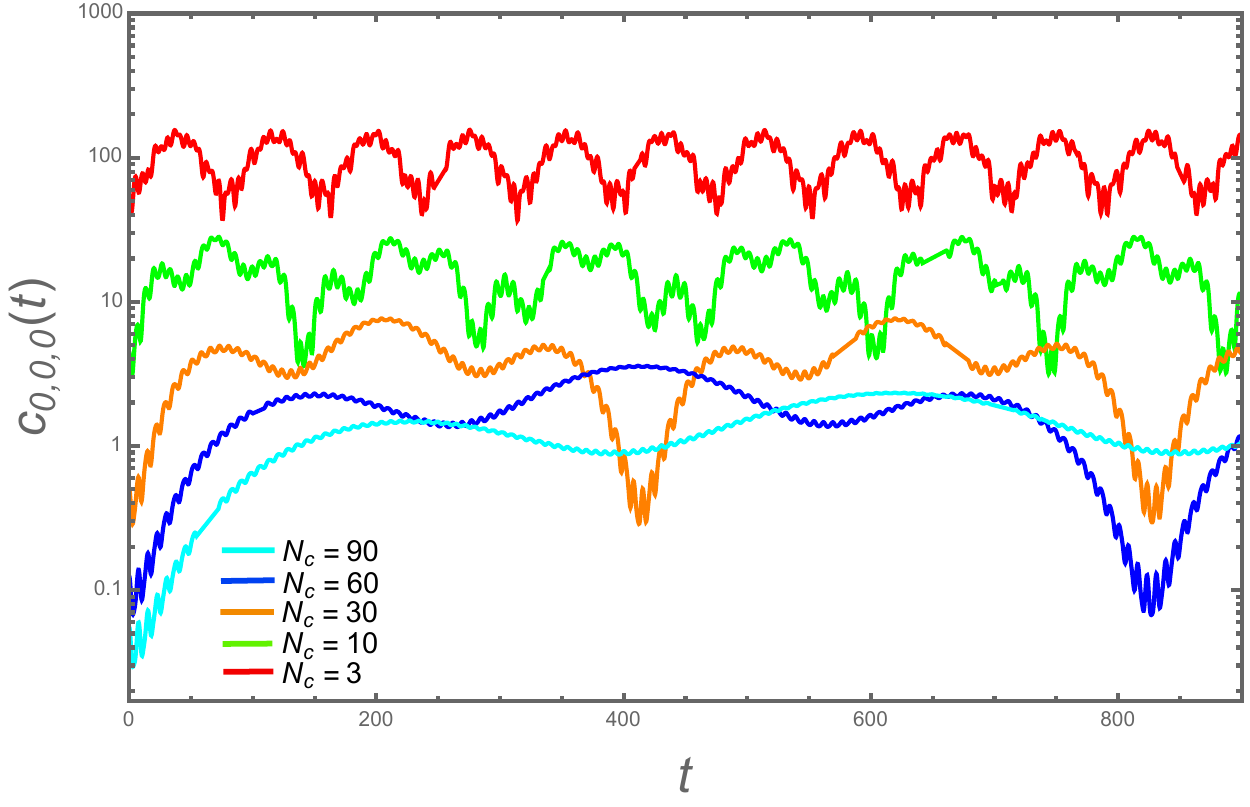}\includegraphics[scale=0.37]{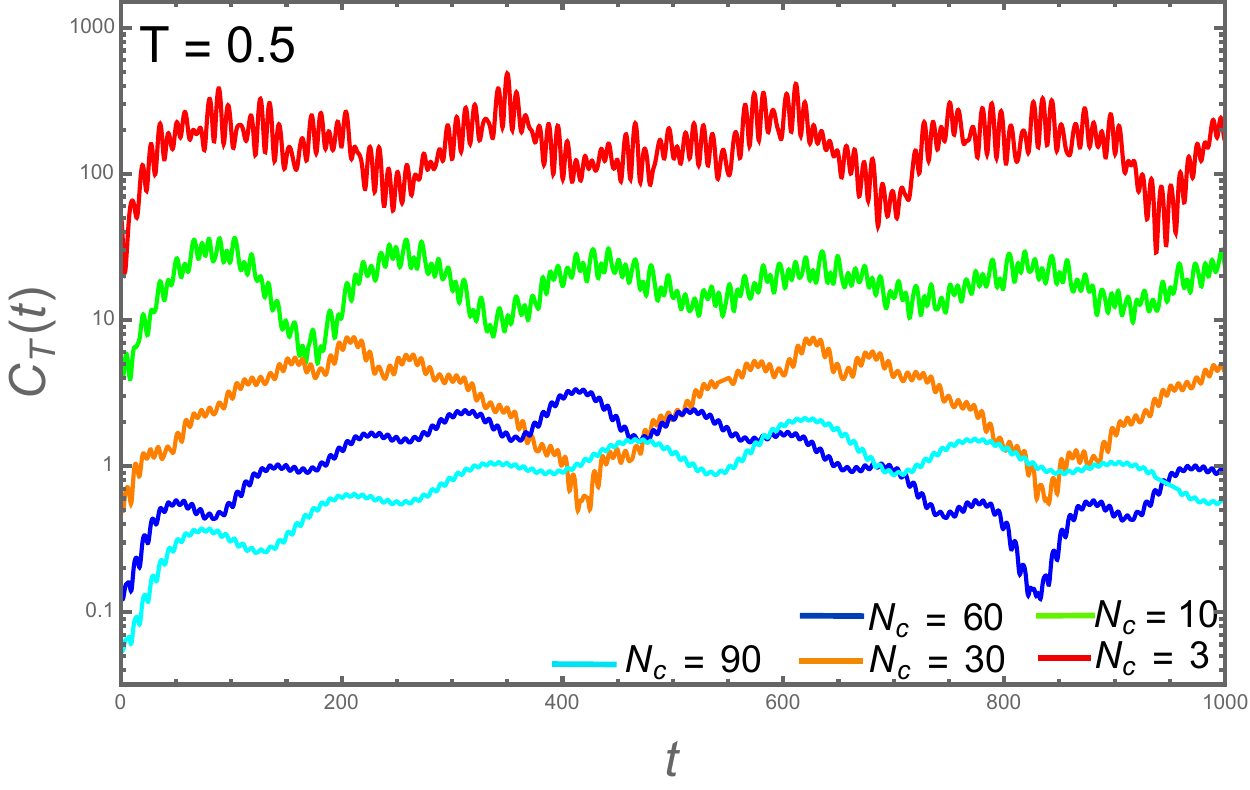}
\par\end{centering}
\begin{centering}
\includegraphics[scale=0.37]{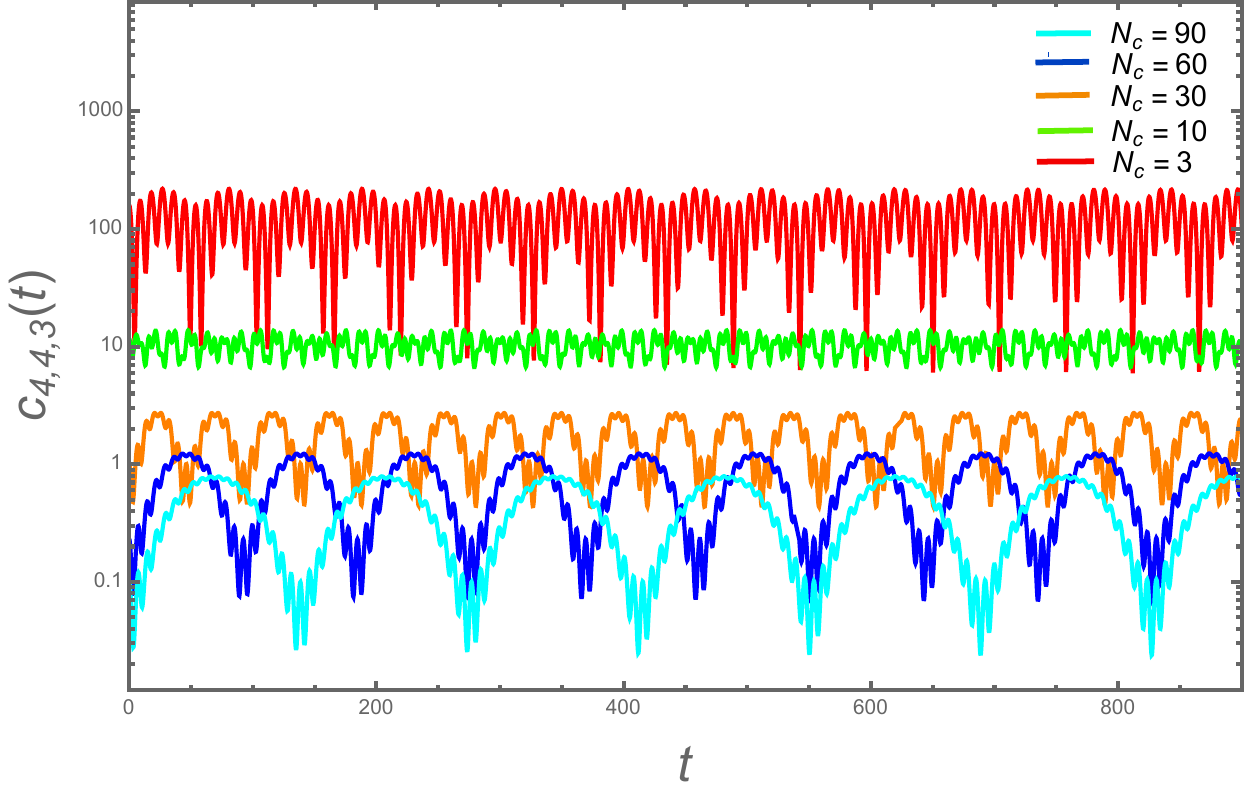}\includegraphics[scale=0.37]{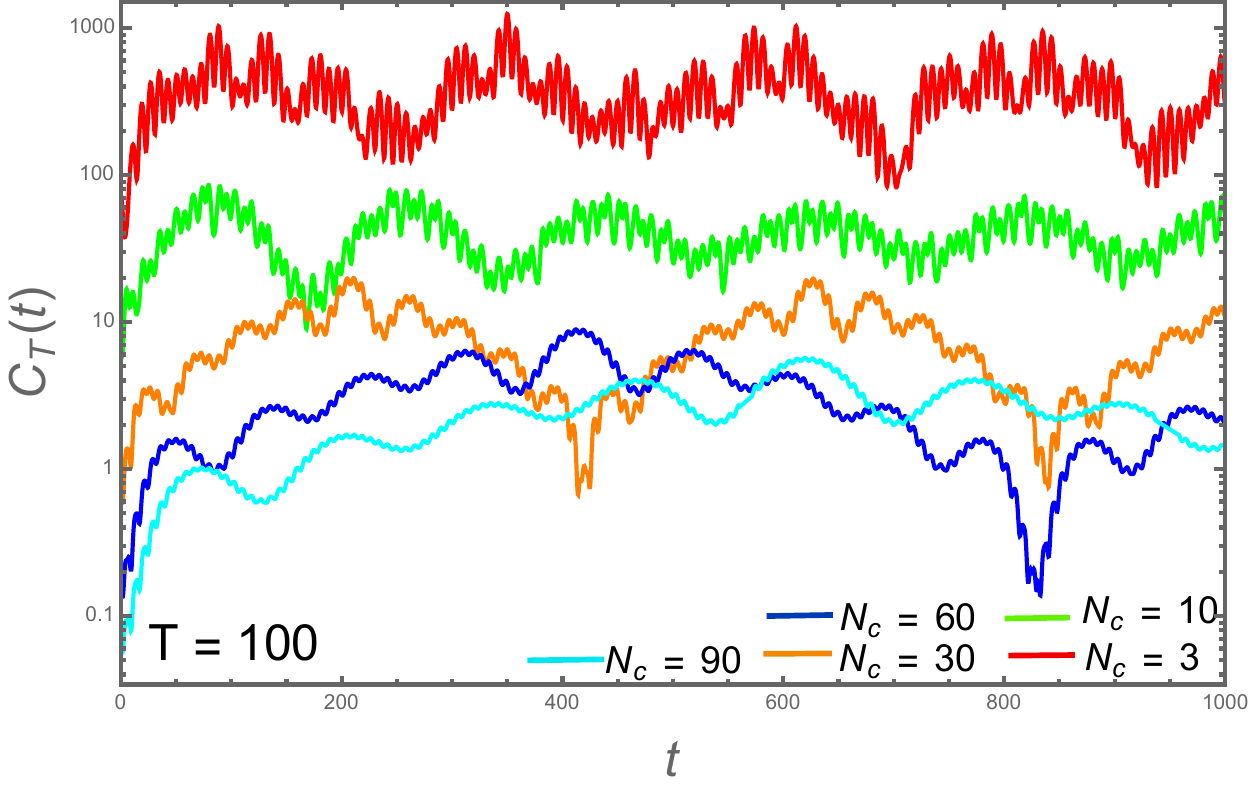}
\par\end{centering}
\caption{\label{fig:2} Microcanonical and thermal OTOC of holographic Skyrmion
with various color number $N_{c}$ and two flavors $N_{f}=2$. \textbf{Left:}
Microcanonical OTOC $c_{0,0,0}\left(t\right)$ and $c_{4,4,3}\left(t\right)$
as functions of time $t$. \textbf{Right:} Thermal OTOC $C_{T}\left(t\right)$
as a function of time $t$ and temperature $T$.}
\end{figure}

 For the two-flavor case i.e. $N_{f}=2$ and $n=N_{f}^{2}-1=3$, we
plot numerically the microcanonical and thermal OTOC of the holographic
Skyrmion with various color numbers $N_{c}$ and temperature $T$
respectively in Figure \ref{fig:1} and \ref{fig:2}. While the formulas
discussed in Section 2 illustrate that microcanonical and thermal
OTOCs are evaluated by the sum of infinite terms, we have truncated
it at $n_{c}=4$ i.e. $0\leq n_{Z},n_{\rho},l_{3}\leq n_{c}$ for
the actual numerical calculation. Notice that (\ref{eq:17}) implies
the energy spectrum only depends on the quantum number $n_{Z},n_{\rho},l_{3}$
for $N_{f}=2$, thus for a given energy (i.e. $n_{Z},n_{\rho},l_{3}$
are all fixed), its degeneracy number $n_{\mathrm{d}}$ can be computed
as $n_{\mathrm{d}}\left(l_{3}\right)=\sum_{l_{2}=0}^{l_{3}}\left(2l_{2}+1\right).$
Therefore, once we truncate the sum for the OTOC given in (\ref{eq:35})
(\ref{eq:37}) and (\ref{eq:38}) at $n_{c}$, it means there are
totally $\left(n_{c}+1\right)^{2}\times\sum_{l_{3}=0}^{n_{c}}n_{\mathrm{d}}\left(l_{3}\right)=1375$
states contributing to the OTOC. However, based on the present derivation,
we have seen the microcanonical OTOC only depends on the quantum numbers
$n_{\rho},l_{3},l_{2}$, so it implies only $\left(n_{c}+1\right)\times\sum_{l_{3}=0}^{n_{c}}\left(l_{3}+1\right)=75$
of the total 1375 OTOCs are independent. Therefore, we denote the
microcanonical OTOC as $c_{n_{Z},n_{\rho},l_{3},l_{2},l_{1}}\left(t\right)\equiv c_{n_{\rho},l_{3},l_{2}}\left(t\right)$
since $n_{Z},l_{1}$ are degenerated quantum number. Besides, we note
that as the excitation states of the Skyrmion given in (\ref{eq:17})
are identified to the baryon states in the D4/D8 model and as it is
known there are no more than 500 types of realistic baryons, so it
is not necessary to take into account the contribution of the whole
baryon spectrum (\ref{eq:17}) to the OTOCs. Hence we chose the truncation
$n_{c}=4$ including 1375 baryon states contributing to the OTOC is
sufficient to cover all the types of realistic baryon. 

\begin{figure}[th]
\begin{centering}
\includegraphics[scale=0.37]{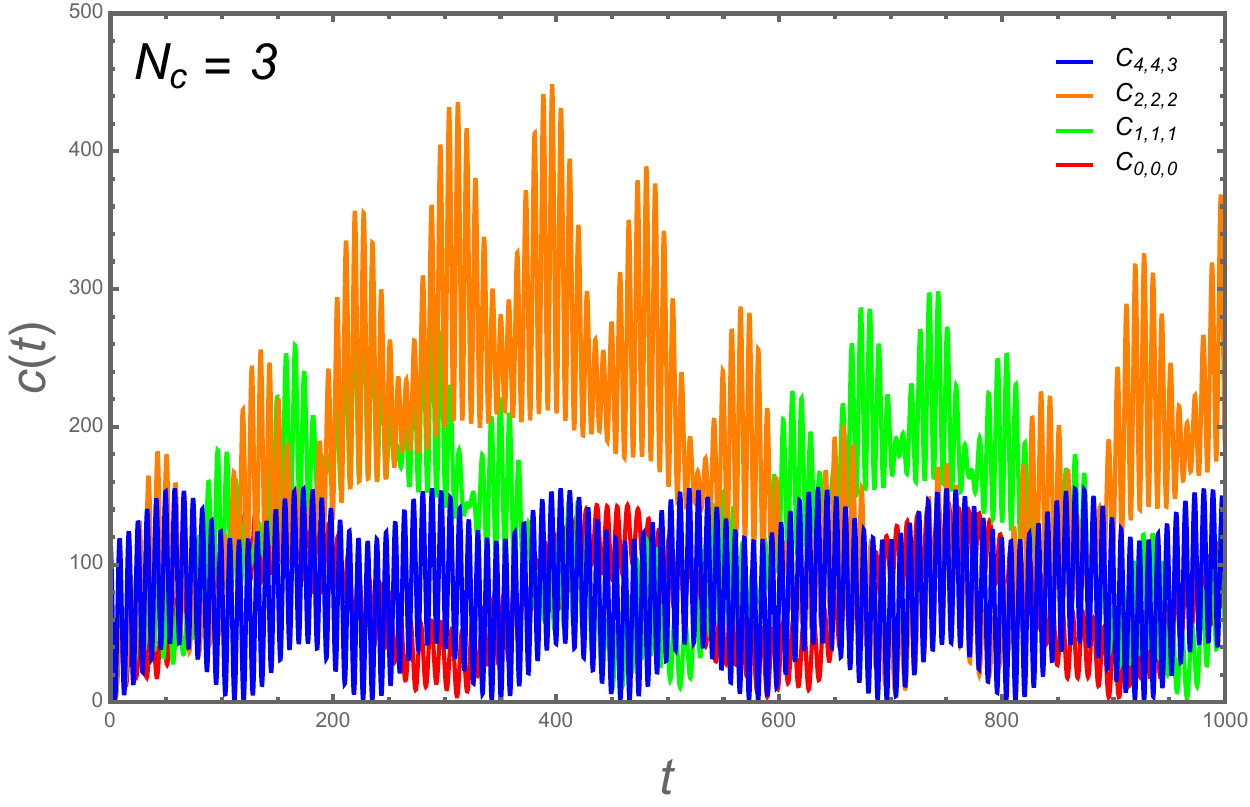}\includegraphics[scale=0.37]{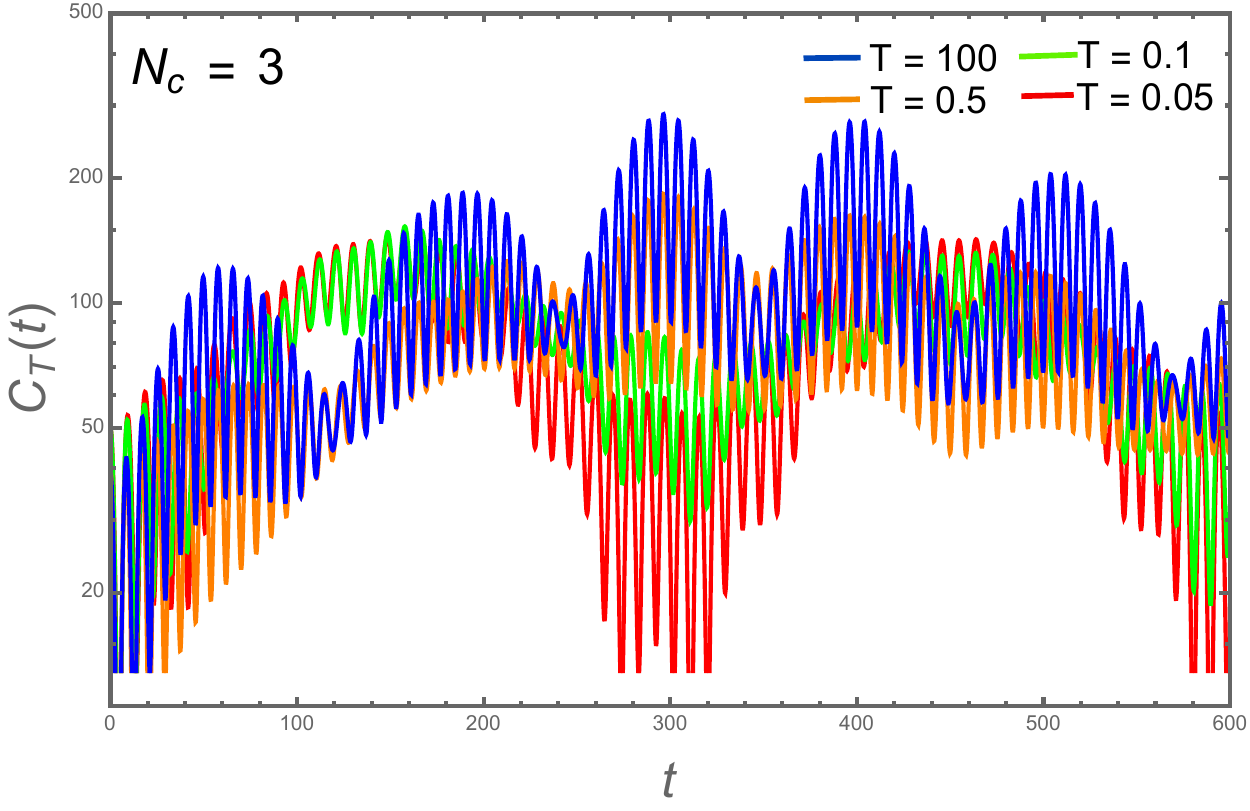}
\par\end{centering}
\begin{centering}
\includegraphics[scale=0.37]{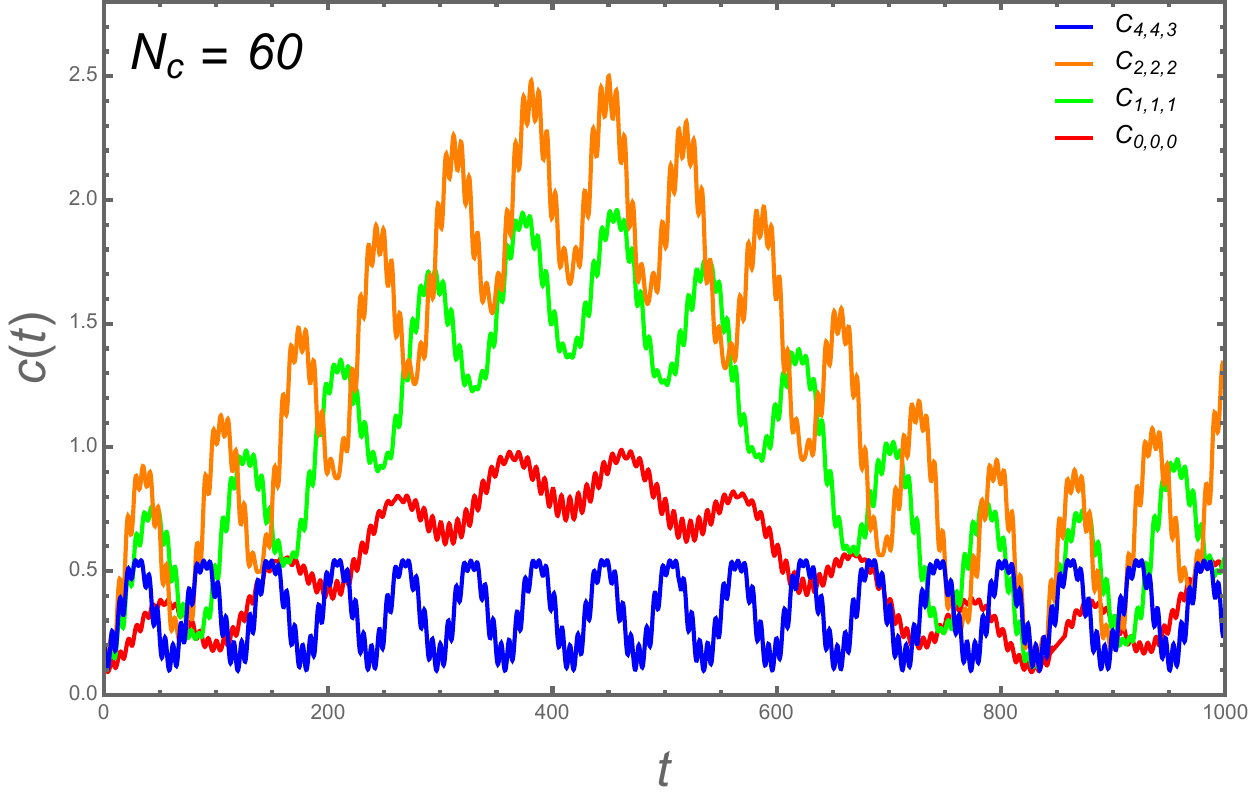}\includegraphics[scale=0.37]{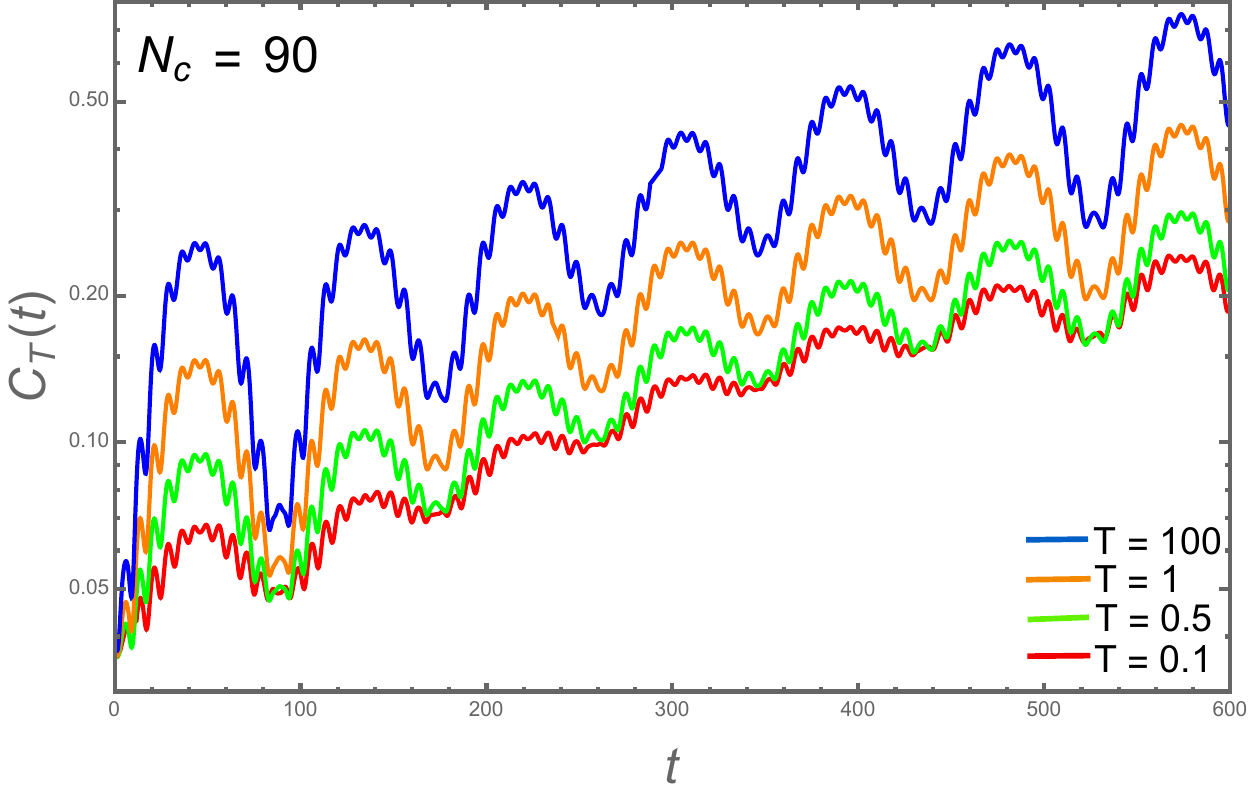}
\par\end{centering}
\caption{\label{fig:3} Microcanonical and thermal OTOC of holographic Skyrmion
with three flavors $N_{f}=3$. \textbf{Left:} Microcanonical OTOC
$c\left(t\right)$ as a function of time $t$. \textbf{Right:} Thermal
OTOC $C_{T}\left(t\right)$ as a function of time $t$ and temperature
$T$.}
\end{figure}
 Keeping these in hands, we can see following features of the OTOC
according to our numerical evaluation. First, the OTOCs grow at early
times while they do not display the apparent exponential growth, at
least for the case $T\leq100$ with $N_{c}\leq90$. Hence, the apparent
exponential growth might probably be the character for a classical
mechanical system distinguishing from a quantum mechanical system
which agrees with \cite{key-3}. Second, the microcanonical OTOCs
are all periodic due to the commensurable energy spectrum. For large
quantum numbers, the microcanonical OTOCs tend to oscillate greatly
because the modes with high frequency become important in (\ref{eq:35}).
Moreover, for the case $N_{c}\gg1$, when we further increase $N_{c}$,
the period of microcanonical OTOCs grows also. This can be clarified
by recalling the energy spectrum (\ref{eq:17}) and the formula (\ref{eq:35})
for the microcanonical OTOC, the period $\Delta t$ of microcanonical
OTOCs can be evaluated as $\Delta t\sim\mathrm{min}\left[E_{n_{\rho},l_{n};n_{\rho}^{\prime},l_{n}^{\prime}}^{-1}\right]\sim N_{c}$
for $N_{c}\rightarrow\infty$ which illustrates the large-$N_{c}$
behaviors of the microcanonical OTOCs. Third, one can see the thermal
OTOCs grow with temperature and additionally tend to become periodic
at early times when $N_{c}$ increases. This is due to the existence
of the sufficiently large period of the microcanonical OTOCs. And,
in any case, even if we write the OTOC as a Lyapunov exponent form
as $C_{T}\sim e^{i\tilde{L}t}$, it illustrates that $\tilde{L}$
vanishes in the large $N_{c}$ limit. Besides, we note that for the
theoretical consistence, it would be less valid to consider the holographic
Skyrmion at very high temperature since the quantum mechanical system
(\ref{eq:9}) comes from the D4 bubble configuration (without a black
hole horizon) which corresponds holographically to a quantum theory
at zero temperature or, at least, at very low temperature \cite{key-19,key-23,key-24}. 

\subsection{Case of three flavors}

\begin{figure}[t]
\begin{centering}
\includegraphics[scale=0.37]{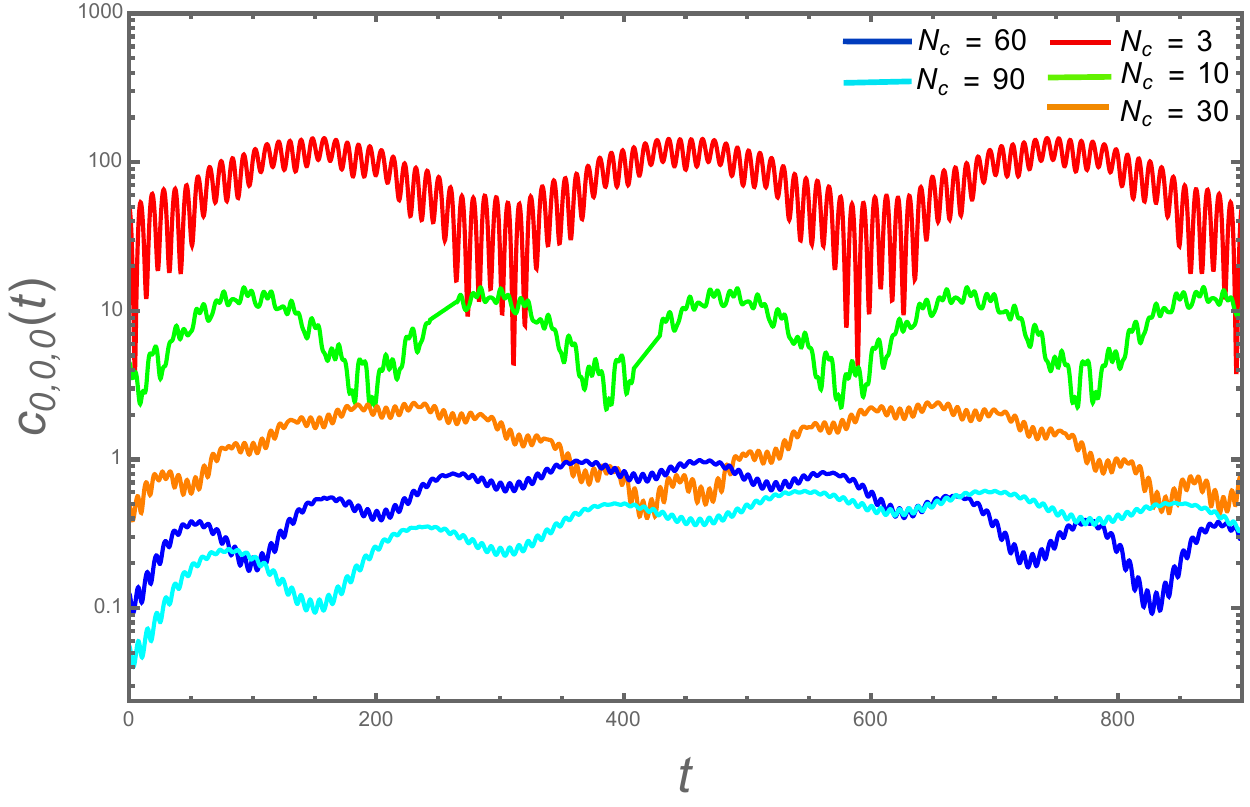}\includegraphics[scale=0.37]{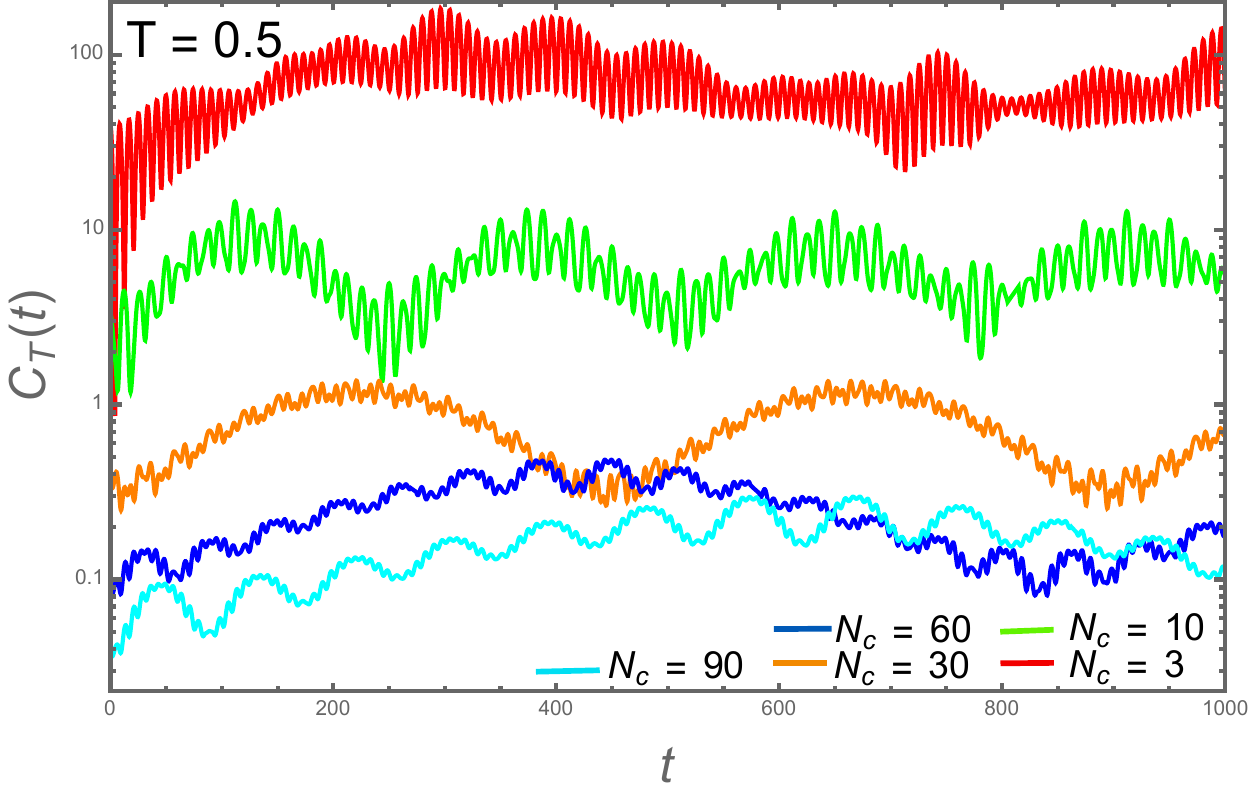}
\par\end{centering}
\begin{centering}
\includegraphics[scale=0.37]{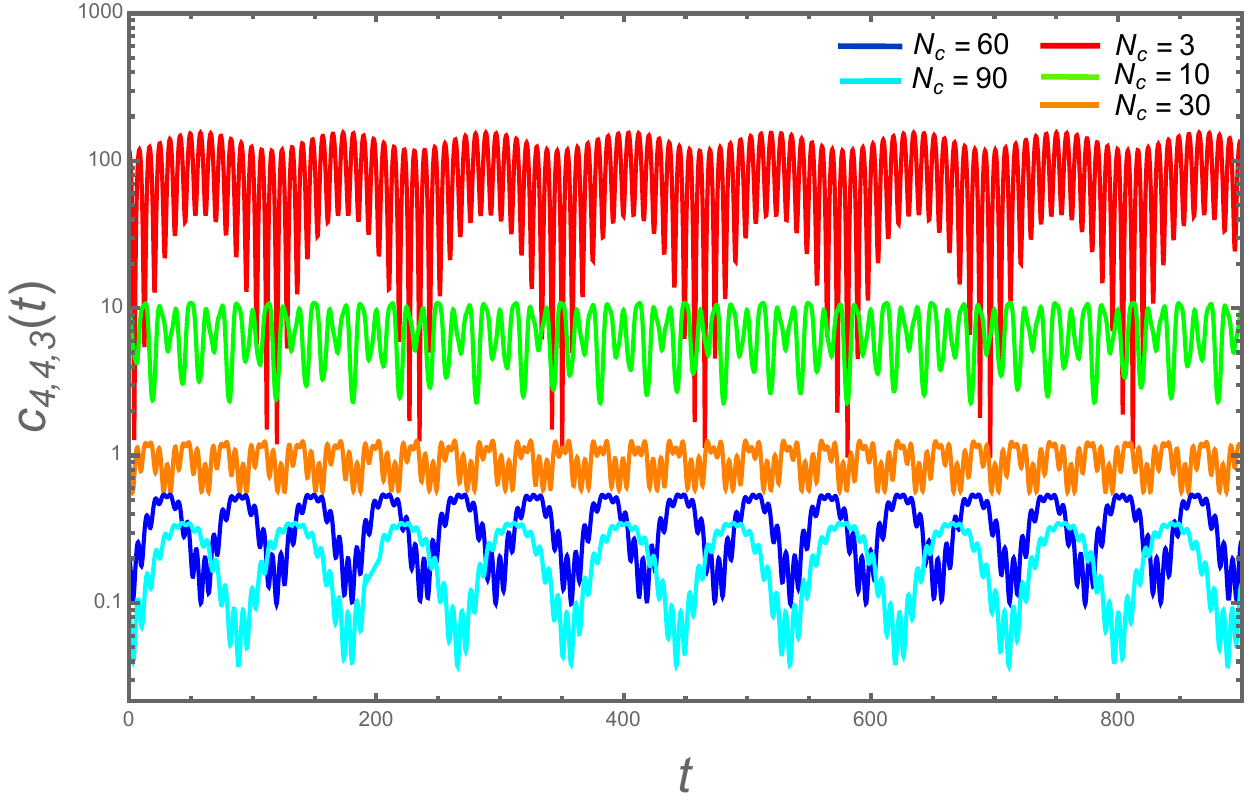}\includegraphics[scale=0.37]{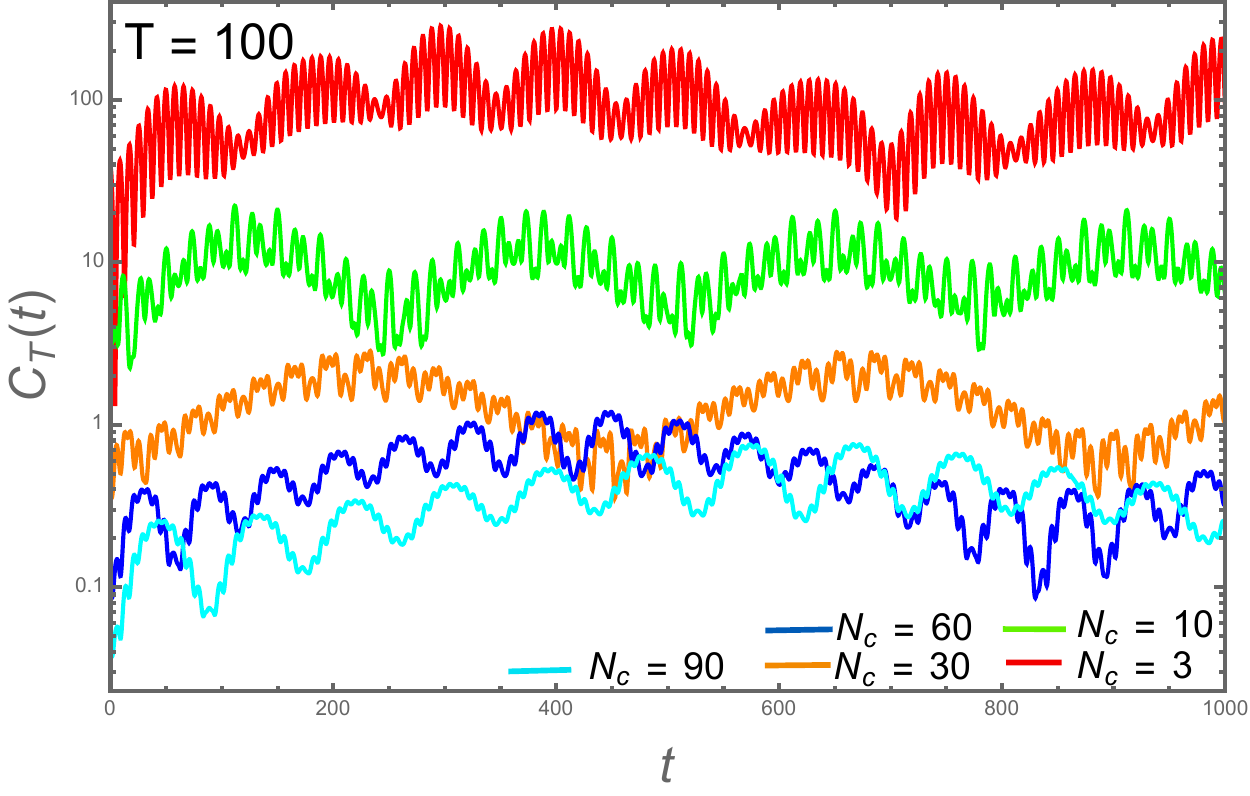}
\par\end{centering}
\caption{\label{fig:4} Microcanonical and thermal OTOC of holographic Skyrmion
with various color number $N_{c}$ and three flavors $N_{f}=3$. \textbf{Left:}
Microcanonical OTOC $c_{0,0,0}\left(t\right)$ and $c_{4,4,3}\left(t\right)$
as functions of time $t$. \textbf{Right:} Thermal OTOC $C_{T}\left(t\right)$
as a function of time $t$ and temperature $T$.}
\end{figure}
 For the three-flavor case i.e. $N_{f}=3$ and $n=N_{f}^{2}-1=8$,
the microcanonical and thermal OTOCs of the holographic Skyrmion with
various color numbers $N_{c}$ and temperature $T$ are numerically
evaluated in Figure \ref{fig:3} and \ref{fig:4}. As the two-flavor
case, we have truncated the sum for OTOCs at $n_{c}=4$ i.e. $0\leq n_{Z},n_{\rho},l_{8}\leq n_{c}$
for the actual numerical calculation. Since the energy spectrum depends
on the quantum number $n_{Z},n_{\rho},l_{8}$ for $N_{f}=3$, in this
case, we can compute the degeneracy number $n_{\mathrm{d}}$ for a
given energy (i.e. $n_{Z},n_{\rho},l_{8}$ are all fixed) which is
given by $n_{\mathrm{d}}\left(l_{8}\right)=\sum_{l_{7}=0}^{l_{8}}...\sum_{l_{3}=0}^{l_{4}}\sum_{l_{2}=0}^{l_{3}}\left(2l_{2}+1\right)$.
Therefore there are totally $\left(n_{c}+1\right)^{2}\times\sum_{l_{8}=0}^{n_{c}}n_{\mathrm{d}}\left(l_{8}\right)=16500$
states contributing to the OTOC at the truncation $n_{c}=4$ while
$\left(n_{c}+1\right)\times\sum_{l_{3}=0}^{n_{c}}\left(l_{3}+1\right)=75$
of the total OTOCs are independent. Accordingly, we denote the microcanonical
OTOC as $c_{n_{Z},n_{\rho},l_{8},...,l_{1}}\left(t\right)\equiv c_{n_{\rho},l_{8},l_{7}}\left(t\right)$
as what we have done in the two-flavor case since $n_{Z},l_{6},...l_{1}$
are degenerated quantum number. Overall, the OTOCs in the three-flavor
case behaves very similarly as those in the two-flavor case because
the growth of the flavor number $N_{f}$ only increases the dimension
of the moduli space as it is discussed in Section 2.1 while the energy
spectrum concerning the OTOC is mostly determined by the radial Hamiltonian
presented in (\ref{eq:9}). 

\section{The classical limit of the OTOC}

In this section, let us study the classical OTOCs of this system by
using the classical statistics. For the microcanonical OTOC, its classical
version can be obtained by naturally replacing the quantum commutators
with Poisson bracket, i.e. $\frac{1}{i}\left[,\right]\rightarrow\left\{ ,\right\} _{\mathrm{P.B}}$.
For the thermal OTOC, we can simply assume that the thermal average
is replaced by the integral in the phase space consisted of canonical
coordinate $\mathbf{q}$ and momentum $\mathbf{p}$. In this sense,
the classical version of the OTOC is given as,

\begin{equation}
C_{T}^{\mathrm{cl}}\left(t\right)=\frac{1}{\mathcal{Z}_{\mathrm{cl}}}\int\frac{d\mathbf{q}d\mathbf{p}}{\left(2\pi\right)^{n+2}}e^{-\beta H}\left\{ \mathbf{q}\left(t\right),\mathbf{p}\left(0\right)\right\} _{\mathrm{P.B}}^{2},c^{\mathrm{cl}}\left(t\right)=\left\{ \mathbf{q}\left(t\right),\mathbf{p}\left(0\right)\right\} _{\mathrm{P.B}}^{2},
\end{equation}
where

\begin{equation}
\mathcal{Z}_{\mathrm{cl}}=\int\frac{d\mathbf{q}d\mathbf{p}}{\left(2\pi\right)^{n+2}}e^{-\beta H}.
\end{equation}
Keeping these in mind, let us investigate the OTOCs of ``classical
Skyrmion'' by using the classical version of the Hamiltonian (\ref{eq:9}),
which is expected to be,

\begin{align}
H= & M_{0}+H_{Z}+H_{y},\nonumber \\
H_{Z}= & \frac{p_{Z}^{2}}{2m_{Z}}+\frac{1}{2}m_{Z}\omega_{Z}^{2}Z^{2},\nonumber \\
H_{y}= & \frac{1}{2m_{\rho}}\sum_{I=1}^{n+1}p_{I}^{2}+\frac{1}{2}m_{\rho}\omega_{\rho}^{2}\rho^{2}+\frac{K}{m_{\rho}\rho^{2}},\label{eq:42}
\end{align}
where $p_{Z},p_{I}$ are the canonical momentum associated to $Z,y_{I}$.
So the canonical coordinate $\mathbf{q}$ and momentum $\mathbf{p}$
is chosen as $\mathbf{q}=\left\{ Z,y_{I}\right\} ,\mathbf{p}=\left\{ p_{Z},p_{I}\right\} $.
For the $Z$ part in (\ref{eq:42}), the classical Hamiltonian $H_{Z}$
reduces to the equation of motion for an 1d harmonic oscillator whose
solution takes the same form as it is given in (\ref{eq:27}). Hence,
according to the solution (\ref{eq:27}), it is very simple to obtain
the classical OTOC as\footnote{The phase space consisted of $\mathbf{q}$ and $\mathbf{p}$ is $2\left(n+2\right)$-dimensional. },

\begin{align}
c_{Z}^{\mathrm{cl}}\left(t\right) & =\left\{ Z\left(t\right),p_{Z}\left(0\right)\right\} _{\mathrm{P.B}}^{2}=\left[\frac{\delta Z\left(t\right)}{\delta Z\left(0\right)}\right]^{2}=\cos^{2}\omega_{Z}t,\nonumber \\
C_{T}^{\mathrm{cl}}\left(t\right) & =\frac{1}{\mathcal{Z}_{\mathrm{cl}}}\int\frac{d\mathbf{q}d\mathbf{p}}{\left(2\pi\right)^{n+2}}e^{-\beta H}\left\{ Z\left(t\right),p_{Z}\left(0\right)\right\} _{\mathrm{P.B}}^{2}=\cos^{2}\omega_{Z}t,
\end{align}
which agrees quantitatively with the quantum OTOC of an 1d harmonic
oscillator as we have discussed in Section 2.2. 

For the $y$ part in (\ref{eq:42}), the classical equation of motion
is given as,

\begin{equation}
m_{\rho}\ddot{y}_{I}=\left(-m_{\rho}\omega_{\rho}^{2}+\frac{2K}{m_{\rho}\rho^{4}}\right)y_{I}.\label{eq:44}
\end{equation}
Since we are considering the classical soliton as the Skyrmion, the
size $\rho$ of the soliton must minimize the soliton energy to be
stable as it is discussed in \cite{key-20,key-21}. Notice the soliton
mass is the potential presented in Hamiltonian (\ref{eq:42}), so
the size of the classical soliton can be evaluated as,

\begin{equation}
\rho^{\mathrm{cl}}=\frac{\left(2K\right)^{1/4}}{\left(m_{\rho}\omega_{\rho}\right)^{1/2}}=3^{7/4}\left(\frac{2}{5}\right)^{1/4}\pi^{1/2},\label{eq:45}
\end{equation}
by minimizing the potential in (\ref{eq:42}). Inserting the size
(\ref{eq:45}) into (\ref{eq:44}), the classical solution for $y_{I}$
is obtained as,

\begin{equation}
y_{I}\left(t\right)=y_{I}\left(0\right)+\frac{p_{I}\left(0\right)}{m_{\rho}}t.
\end{equation}
In this sense, the classical OTOC is evaluated as,

\begin{align}
c_{y_{I}}^{\mathrm{cl}} & =\left\{ y_{I}\left(t\right),p_{I}\left(0\right)\right\} _{\mathrm{P.B}}^{2}=1,\nonumber \\
C_{T}^{\mathrm{cl}}\left(t\right) & =\frac{1}{\mathcal{Z}_{\mathrm{cl}}}\int\frac{d\mathbf{q}d\mathbf{p}}{\left(2\pi\right)^{n+2}}e^{-\beta H}\left\{ y_{I}\left(t\right),p_{I}\left(0\right)\right\} _{\mathrm{P.B}}^{2}=1,
\end{align}
which means the chaos does not evolve. While this result is apparently
different from the quantum OTOC, it is reasonable because the classical
soliton described by the BPST solution is static. In addition if we
take (\ref{eq:45}), it implies $\rho$ could not be a collective
coordinate i.e. time-independent. While this agrees with some discussions
about Skyrmion i.e. \cite{key-43,key-44,key-45}, our current holographic
approach supports the opposite \cite{key-20,key-21}. Without loss
of generality, let us consider a small correction $d$ to the size
of the classical soliton i.e. 
\begin{equation}
\rho=\rho^{\mathrm{cl}}\left(1-d\right),\left|d\right|\ll1,
\end{equation}
which means the soliton is metastable. In this case, we can obtain
the solution for the classical equation of motion (\ref{eq:44}) as,

\begin{equation}
y_{I}\left(t\right)=y_{I}\left(0\right)e^{Lt},\ L=\frac{1}{\left(1-d\right)^{2}}\sqrt{\frac{d\left(4-6d+4d^{2}-d^{3}\right)}{6}}\simeq\sqrt{\frac{2d}{3}}+\mathcal{O}\left(d^{2}\right).
\end{equation}
Therefore, the classical OTOC is evaluated as,

\begin{equation}
c_{y_{I}}^{\mathrm{cl}},C_{T}^{\mathrm{cl}}\propto e^{2Lt},\ \left|L\right|\rightarrow0,
\end{equation}
which means they would grow exponentially if $d>0$, or oscillate
if $d<0$. So we can find the classical OTOCs are very sensitive to
the size of the soliton. In that way, if we consider a soliton with
size $\rho$ as the initial condition, $\rho$ will become to $\rho^{\mathrm{cl}}$
rapidly. That implies the OTOC is exponential at very early then trends
to be a constant quickly. Interestingly, while this is the interpretation
of the behaviors of the classical OTOCs, it very roughly covers the
behaviors of the quantum thermal OTOCs at large $t$ as well.

\section{Summary and discussion}

In this work, we study the OTOC of the holographic Skyrmion as baryon
in the D4/D8 model. Since the quantum mechanical system for the holographic
Skyrmion is totally analytical, the derivation of the OTOC is based
on the standard technique in quantum mechanics. Our numerical evaluation
illustrates the following outcomes of the OTOC. First, in any case
we do not see the exponential growth in the quantum OTOC while it
can be present in the classical OTOC, thus it supports the conjecture
that the exponential growth may be the feature of the classical OTOC
only. Second, the thermal OTOCs tend to become periodic when $N_{c}$
increases. This behavior implies the Lyapunov exponent $L$ of this
system is vanished in the large $N_{c}$ limit, and this conclusion
remains even if we define the imaginary Lyapunov exponent $\tilde{L}$
as $C_{T}\sim e^{i\tilde{L}t}$ since we have shown that $\tilde{L}\sim N_{c}^{-1}$.
Therefore, this large $N_{c}$ behavior of the Lyapunov exponent agrees
with \cite{key-4} and the behavior of the classical OTOC discussed
in Section 4. Remarkably, it might be interesting, because if the
Lyapunov exponent saturates, according to AdS/CFT, the dual quantum
theory may be able to describe some features of the bulk black hole
horizon holographically. So think the opposite, it means if the Lyapunov
exponent vanishes, the dual quantum theory should not describe the
bulk black hole horizon. This analysis is seemingly valid because
both the bulk geometries of the D4/D8 and D3/D7 used in \cite{key-4}
do not have an event horizon in the view of holography. Third, there
could be a time scale that the quantum OTOCs trend to saturate which
may be the boundary of the distinction between particle-like and wave-like
behavior of the OTOC in this system. And the trend of the OTOC is
not very sensitive to the truncation if the sufficiently enough states
are included into the calculation. We also study the OTOC with 3d
Coulomb potential to support these analyses.

Besides, let us give some more comments to close this work. The OTOC
may be a measure to test the validation of the holographic correspondence
between a gravity and a quantum theory. However in the QT (quantum
theory) side, only few models display the exponential growth in OTOC,
e.g. SYK model \cite{key-14,key-15}, hence the essential for the
exponential growth in OTOC is still less clear. A possible way to
explore this issue in holography may be to construct a holographic
quantum mechanics or quantum field theory by using the bulk geometry
with black hole, then compute the OTOC in the dual theory to find
the essential for the exponential growth. While this may be instructive,
the dual quantum theory could be very complex to compute the OTOC
since it could be thermal field theory or quantum many-body systems.
Therefore, we will attempt to find a holographic quantum mechanical
system simple enough and corresponding to some bulk geometries with
an event horizon in order to figure out these issues in future works.

\section*{Acknowledgements}

This work is supported by the National Natural Science Foundation
of China (NSFC) under Grant No. 12005033 and the Fundamental Research
Funds for the Central Universities under Grant No. 3132023198.

\section{Appendix A: Spherical harmonic function on $S^{n}$ and the useful
matrix elements}

In this appendix, we first construct the spherical harmonic function
in arbitrary dimension, then derive the kernel function $Y_{l_{n}^{\prime},l_{n-1}^{\prime};l_{n},l_{n-1}}^{\mathrm{ker}}$
as the key part of the matrix element present in Section 2.2 for the
OTOCs. Consider the $n+1$ dimensional Euclidean space $\mathbb{R}^{n+1}$
parametrized by $x^{i}=\left\{ x^{1},x^{2},...x^{n+1}\right\} $ as
the Cartesian coordinates. Turning to the spherical coordinates $\left\{ \rho,\theta_{1},...,\theta_{n-1},\theta_{n}\right\} $,
we have the following coordinate transformation

\[
x^{i}=\rho\cos\theta_{i-1}\prod_{j=i}^{n}\sin\theta_{j},\ \rho^{2}=\sum_{i=1}^{n+1}\left(x^{i}\right)^{2},\tag{A-1}
\]
where $\theta_{0}=0$. The Euclidean metric on $\mathbb{R}^{n+1}$
is given as,

\[
ds^{2}=\delta_{ij}dx^{i}dx^{j}=d\rho^{2}+\rho^{2}d\Omega_{n}^{2},\tag{A-2}
\]
where $d\Omega_{n}^{2}$ refers to the angular differential on a unit
$S^{n}$ satisfying the recurrence as,

\[
d\Omega_{n}^{2}=d\theta_{n}^{2}+\sin^{2}\theta_{n}d\Omega_{n-1}^{2}.\tag{A-3}
\]
Hence we can obtain,

\[
d\Omega_{n}^{2}=\sum_{i=1}^{n}\left(d\theta_{i}^{2}\prod_{j=i+1}^{n}\sin^{2}\theta_{j}\right),\tag{A-4}
\]
and the volume element $dV_{n+1}$ of $\mathbb{R}^{n+1}$ represented
by the differentiation of spherical coordinates reads as

\begin{align*}
dV_{n+1} & =\sqrt{g}d\rho d\theta_{n}d\theta_{n-1}...d\theta_{1}=\rho^{n}dr\prod_{j=1}^{n}\sin^{j-1}\theta_{j}d\theta_{j}\equiv\rho^{n}drdV_{S^{n}}.\tag{A-5}
\end{align*}
The spherical harmonic function $\mathcal{Y}_{l_{n},l_{n-1},...l_{1}}\left(\theta_{n},\theta_{n-1},...\theta_{1}\right)$
on $S^{n}$ is the eigenfunction of Laplacian operator $\nabla_{S^{n}}^{2}$
satisfying

\[
\nabla_{S^{n}}^{2}\mathcal{Y}_{l_{n},l_{n-1},...l_{1}}\left(\theta_{n},\theta_{n-1},...\theta_{1}\right)=-l_{n}\left(l_{n}+n-1\right)\mathcal{Y}_{l_{n},l_{n-1},...l_{1}}\left(\theta_{n},\theta_{n-1},...\theta_{1}\right),\tag{A-6}
\]
where the quantum number satisfies $\left|l_{1}\right|\leq l_{2}\leq l_{3}\leq...l_{n}.$
Note that the Laplacian operator $\nabla_{S^{n}}^{2}$ satisfies the
recurrence as,

\[
\nabla_{S^{n+1}}^{2}=\sin^{-n}\theta_{n+1}\frac{\partial}{\partial\theta_{n+1}}\sin^{n}\theta_{n+1}\frac{\partial}{\partial\theta_{n+1}}+\sin^{-2}\theta_{n+1}\nabla_{S^{n}}^{2},\tag{A-7}
\]
where $\theta_{n+1}$ is the $n+1$-th spherical coordinate on $S^{n+1}$.
Thus the exact formula for $\mathcal{Y}_{l_{n},l_{n-1},...l_{1}}$
can be constructed by using the polynomial $\ _{j}\mathcal{P}_{l_{j}}^{l_{j-1}}\left(\theta_{j}\right)$
as,

\begin{align}
\mathcal{Y}_{l_{n},l_{n-1},...l_{1}}\left(\theta_{n},\theta_{n-1},...\theta_{1}\right) & =\frac{\left(-1\right)^{l_{1}}}{\sqrt{2\pi}}e^{il_{1}\theta_{1}}\prod_{j=2}^{n}\ _{j}\mathcal{P}_{l_{j}}^{l_{j-1}}\left(\theta_{j}\right),\tag{A-8}\label{eq:A-8}
\end{align}
where $\ _{j}\mathcal{P}_{l_{j}}^{l_{j-1}}\left(\theta_{j}\right)$
is given by combining the associated Legendre polynomial $P_{l}^{m}\left(\cos\theta\right)$
as,

\begin{equation}
\ \ \ \ \ \ \ \ \ \ \ \ \ _{j}\mathcal{P}_{l_{j}}^{l_{j-1}}\left(\theta_{j}\right)=\sqrt{\frac{2l_{j}+j-1}{2}\frac{\left(l_{j}+l_{j-1}+j-2\right)!}{\left(l_{j}-l_{j-1}\right)!}}\sin^{\frac{2-j}{2}}\theta_{j}P_{l_{j}+\frac{j-2}{2}}^{-\left(l_{j-1}+\frac{j-2}{2}\right)}\left(\cos\theta_{j}\right).\tag{A-9}\label{eq:A-9}
\end{equation}
The spherical harmonic function $\mathcal{Y}_{l_{n},l_{n-1},...l_{1}}\left(\theta_{n},\theta_{n-1},...\theta_{1}\right)$
also satisfies the normalization condition as

\begin{equation}
\int\mathcal{Y}_{l_{n}^{\prime},l_{n-1}^{\prime},...l_{1}^{\prime}}^{*}\mathcal{Y}_{l_{n},l_{n-1},...l_{1}}dV_{S^{n}}=\prod_{j=1}^{n}\delta_{l_{j}^{\prime}l_{j}}.\tag{A-10}
\end{equation}
Note that the associated Legendre polynomial $P_{l}^{m}\left(\cos\theta\right)$
satisfies the following identity,

\begin{equation}
\frac{l+m}{2l+1}P_{l-1}^{m}\left(\cos\theta\right)+\frac{l-m+1}{2l+1}P_{l+1}^{m}\left(\cos\theta\right)=\cos\theta P_{l}^{m}\left(\cos\theta\right).\tag{A-11}\label{eq:A-11}
\end{equation}
Keep this in hand, let us see the relation of $\cos\theta_{n}\mathcal{Y}_{l_{n},l_{n-1},...l_{1}}$
and $\mathcal{Y}_{l_{n},l_{n-1},...l_{1}}$ by computing

\begin{align}
\cos\theta_{n}\mathcal{Y}_{l_{n},l_{n-1},...l_{1}} & =\frac{\left(-1\right)^{l_{1}}}{\sqrt{2\pi}}e^{il_{1}\theta_{1}}\cos\theta_{n}\prod_{j=2}^{n}\ _{j}\mathcal{P}_{l_{j}}^{l_{j-1}}\left(\theta_{j}\right)\nonumber \\
 & =\frac{\left(-1\right)^{l_{1}}}{\sqrt{2\pi}}e^{il_{1}\theta_{1}}\cos\theta_{n}\ _{2}\mathcal{P}_{l_{2}}^{l_{1}}\left(\theta_{2}\right)\ _{3}\mathcal{P}_{l_{3}}^{l_{2}}\left(\theta_{3}\right)...\ _{n-1}\mathcal{P}_{l_{n-1}}^{l_{n-2}}\left(\theta_{n-1}\right)\ _{n}\mathcal{P}_{l_{n}}^{l_{n-1}}\left(\theta_{n}\right).\tag{A-12}\label{eq:A-12}
\end{align}
The relevant part is given as

\begin{equation}
\cos\theta_{n}\times\ _{n}\mathcal{P}_{l_{n}}^{l_{n-1}}\left(\theta_{n}\right)=\sqrt{\frac{2l_{n}+n-1}{2}\frac{\left(l_{n}+l_{n-1}+n-2\right)!}{\left(l_{n}-l_{n-1}\right)!}}\sin^{\frac{2-n}{2}}\theta_{n}\cos\theta_{n}P_{l_{n}+\frac{n-2}{2}}^{-\left(l_{n-1}+\frac{n-2}{2}\right)}\left(\cos\theta_{n}\right).\tag{A-13}
\end{equation}
Then use relation (\ref{eq:A-11}), we find

\begin{align}
\cos\theta_{n}P_{l_{n}+\frac{n-2}{2}}^{-\left(l_{n-1}+\frac{n-2}{2}\right)}\left(\cos\theta_{n}\right)= & \frac{l_{n}-l_{n-1}}{2l_{n}+n-1}P_{l_{n}+\frac{n-2}{2}-1}^{-\left(l_{n-1}+\frac{n-2}{2}\right)}\left(\cos\theta_{n}\right)\nonumber \\
 & +\frac{l_{n}+l_{n-1}+n-1}{2l_{n}+n-1}P_{l_{n}+\frac{n-2}{2}+1}^{-\left(l_{n-1}+\frac{n-2}{2}\right)}\left(\cos\theta_{n}\right).\tag{A-14}
\end{align}
On the other hand, recall the definition (\ref{eq:A-8}) (\ref{eq:A-9}),
we have,

\begin{equation}
\mathcal{Y}_{l_{n}\pm1,l_{n-1},...l_{1}}\left(\theta_{n},\theta_{n-1},...\theta_{1}\right)=\frac{\left(-1\right)^{l_{1}}}{\sqrt{2\pi}}e^{il_{1}\theta_{1}}\ _{2}\mathcal{P}_{l_{2}}^{l_{1}}\left(\theta_{2}\right)\ _{3}\mathcal{P}_{l_{3}}^{l_{2}}\left(\theta_{3}\right)...\ _{n-1}\mathcal{P}_{l_{n-1}}^{l_{n-2}}\left(\theta_{n-1}\right)\ _{n}\mathcal{P}_{l_{n}\pm1}^{l_{n-1}}\left(\theta_{n}\right),\tag{A-15}
\end{equation}
and

\begin{align}
\ _{n}\mathcal{P}_{l_{n}+1}^{l_{n-1}}\left(\theta_{n}\right) & =\sqrt{\frac{2l_{n}+n+1}{2}\frac{\left(l_{n}+l_{n-1}+n-1\right)!}{\left(l_{n}-l_{n-1}+1\right)!}}\sin^{\frac{2-n}{2}}\theta_{n}P_{l_{n}+\frac{n-2}{2}+1}^{-\left(l_{n-1}+\frac{n-2}{2}\right)}\left(\cos\theta_{n}\right),\nonumber \\
\ _{n}\mathcal{P}_{l_{n}-1}^{l_{n-1}}\left(\theta_{n}\right) & =\sqrt{\frac{2l_{n}+n-3}{2}\frac{\left(l_{n}+l_{n-1}+n-3\right)!}{\left(l_{n}-l_{n-1}-1\right)!}}\sin^{\frac{2-n}{2}}\theta_{n}P_{l_{n}+\frac{n-2}{2}-1}^{-\left(l_{n-1}+\frac{n-2}{2}\right)}\left(\cos\theta_{n}\right).\tag{A-16}
\end{align}
So pick up all the above, we can reach,

\begin{align}
\cos\theta_{n}\ _{n}\mathcal{P}_{l_{n}}^{l_{n-1}}\left(\theta_{n}\right)= & \sqrt{\frac{\left(l_{n}-l_{n-1}\right)\left(l_{n}+l_{n-1}+n-2\right)}{\left(2l_{n}+n-1\right)\left(2l_{n}+n-3\right)}}\ _{n}\mathcal{P}_{l_{n}-1}^{l_{n-1}}\left(\theta_{n}\right)\nonumber \\
 & +\sqrt{\frac{\left(l_{n}+l_{n-1}+n-1\right)}{\left(2l_{n}+n-1\right)}\frac{\left(l_{n}-l_{n-1}+1\right)}{\left(2l_{n}+n+1\right)}}\ _{n}\mathcal{P}_{l_{n}+1}^{l_{n-1}}\left(\theta_{n}\right).\tag{A-17}\label{eq:A-17}
\end{align}
Inserting (\ref{eq:A-17}) into (\ref{eq:A-12}), we finally have

\begin{align*}
\cos\theta_{n}\mathcal{Y}_{l_{n},l_{n-1},...l_{1}}= & \sqrt{\frac{\left(l_{n}-l_{n-1}\right)\left(l_{n}+l_{n-1}+n-2\right)}{\left(2l_{n}+n-1\right)\left(2l_{n}+n-3\right)}}\mathcal{Y}_{l_{n}-1,l_{n-1},...l_{1}}\\
 & +\sqrt{\frac{\left(l_{n}+l_{n-1}+n-1\right)}{\left(2l_{n}+n-1\right)}\frac{\left(l_{n}-l_{n-1}+1\right)}{\left(2l_{n}+n+1\right)}}\mathcal{Y}_{l_{n}+1,l_{n-1},...l_{1}},\tag{A-18}
\end{align*}
which reduces to the useful integration as,

\begin{align}
Y_{l_{n}^{\prime},l_{n-1}^{\prime},...l_{1}^{\prime};l_{n},l_{n-1},...l_{1}}\equiv & \int dV_{S^{n}}\mathcal{Y}_{l_{n}^{\prime},l_{n-1}^{\prime},...l_{1}^{\prime}}^{*}\cos\theta_{n}\mathcal{Y}_{l_{n},l_{n-1},...l_{1}}\nonumber \\
= & \bigg[\sqrt{\frac{\left(l_{n}-l_{n-1}\right)\left(l_{n}+l_{n-1}+n-2\right)}{\left(2l_{n}+n-1\right)\left(2l_{n}+n-3\right)}}\delta_{l_{n}^{\prime},l_{n}-1}\nonumber \\
 & +\sqrt{\frac{\left(l_{n}+l_{n-1}+n-1\right)}{\left(2l_{n}+n-1\right)}\frac{\left(l_{n}-l_{n-1}+1\right)}{\left(2l_{n}+n+1\right)}}\delta_{l_{n}^{\prime},l_{n}+1}\bigg]\times\prod_{j=1}^{n-1}\delta_{l_{j}^{\prime},l_{j}}.\tag{A-19}\label{eq:A-19}
\end{align}
Therefore it reads the kernel function $Y_{l_{n}^{\prime},l_{n-1}^{\prime};l_{n},l_{n-1}}^{\mathrm{ker}}$
as it is given in (\ref{eq:2.33}).

\section{Appendix B: The OTOC in quantum mechanics with 3d Coulomb potential}

Recall the formulas presented in (\ref{eq:19}), we can see the concerned
OTOC is defined as,

\[
C_{T}\left(t\right)=\frac{1}{\mathcal{Z}}\sum_{n}e^{-\beta E_{n}}c_{n}\left(t\right),\ \mathcal{Z}=\sum_{n}e^{-\beta E_{n}},\ c_{n}\left(t\right)=-\left\langle n\left|\left[x_{a}\left(t\right),p_{a}\right]^{2}\right|n\right\rangle ,\tag{B-1}
\]
where the index $a$ refers to the $a$-th direction of the space.
Due to the rotation symmetry, it is easy to verify the classical OTOC
defined in (\ref{eq:19}) is invariant under the $SO\left(N\right)$
rotation, however this conclusion may not remain if we consider its
quantum version. Thus, in this appendix, we discuss briefly the quantum
OTOC with respect to different directions with 3d Coulomb potential,
as another central force field parallel to the soliton potential in
the main text for simplification.

In quantum mechanics, the eigen equation of 3d Hamiltonian with Coulomb
potential is given as,

\[
H\psi_{n,l,m}\left(r,\theta,\phi\right)=E_{n}\psi_{n,l,m},\tag{B-2}
\]
where the Hamiltonian is 
\begin{figure}[t]
\begin{centering}
\includegraphics[scale=0.37]{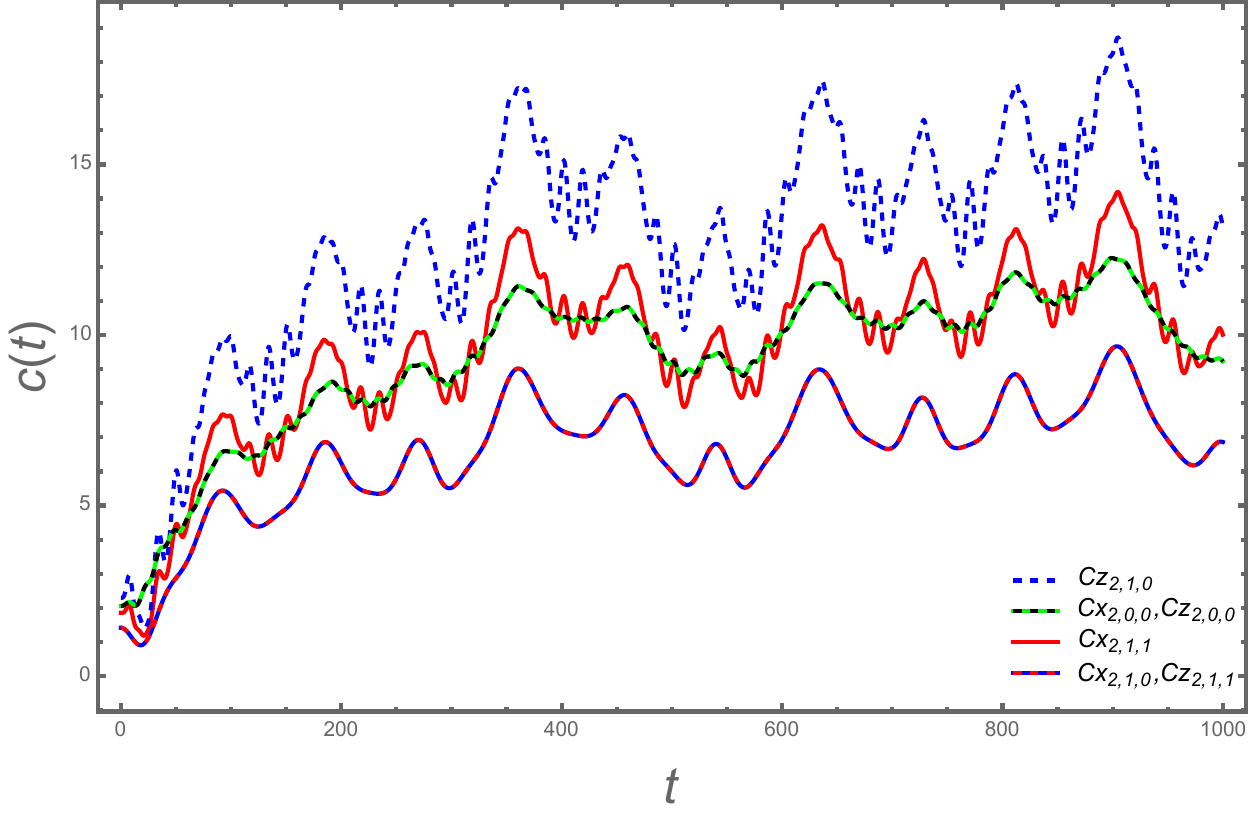}\includegraphics[scale=0.37]{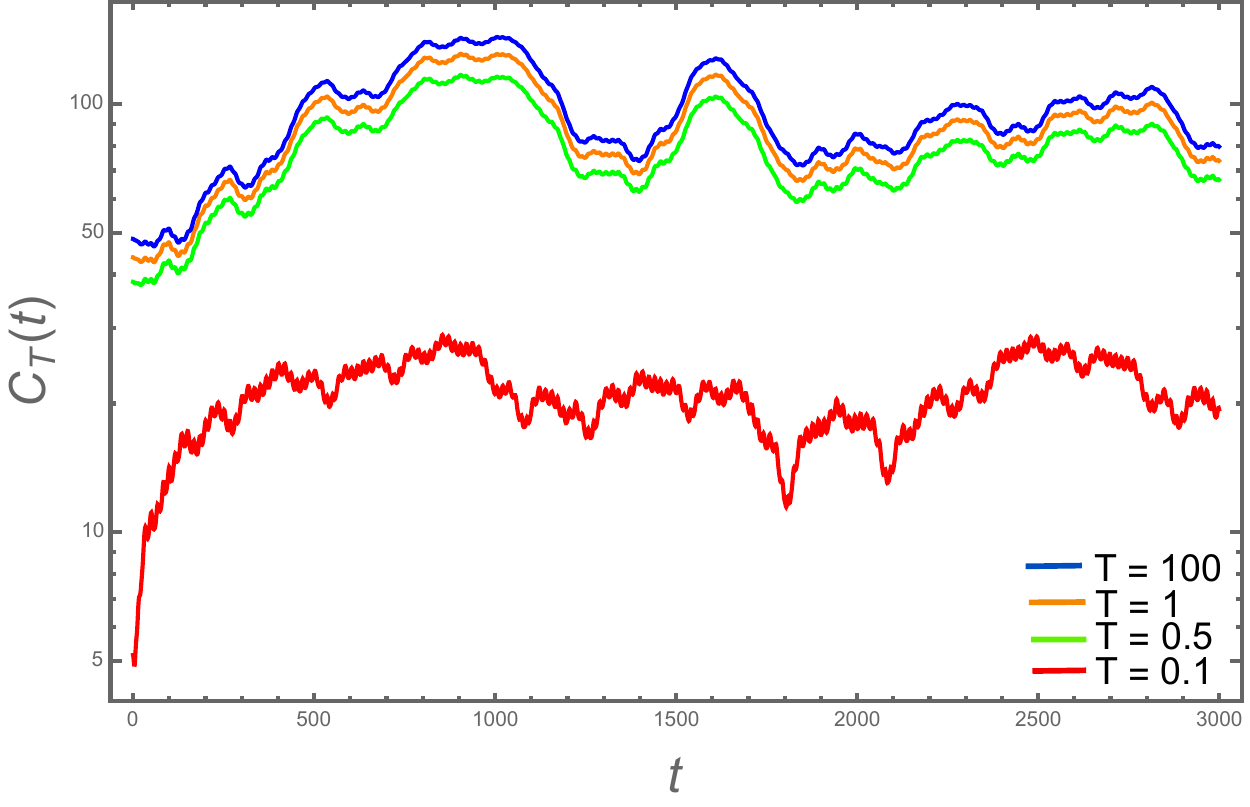}
\par\end{centering}
\caption{\label{fig:B} The OTOCs in quantum mechanics with 3d Coulomb potential
in the unit of $m_{e}=e=1$. \textbf{Left: }The microcanonical OTOCs
$c_{n}^{X}$ and $c_{n}^{Z}$. \textbf{Right:} The thermal OTOC $C_{T}^{X}$
and $C_{T}^{Z}$. Note that, for a fixed temperature, $C_{T}^{X}$
exactly covers $C_{T}^{Z}$ which therefore is not plotted in the
figure and the truncation $n_{c}$ is chosen as $n\protect\leq n_{c},n_{c}=7$.}
\end{figure}

\[
H=-\frac{1}{2m_{e}}\left[\frac{1}{r^{2}}\frac{\partial}{\partial r}\left(r^{2}\frac{\partial}{\partial r}\right)+\frac{1}{r^{2}}\nabla_{S^{2}}^{2}\right]-\frac{e^{2}}{r}.\tag{B-3}
\]
Note that the 3d Euclidean space is parametrized by the spherical
coordinates $\left\{ r,\theta,\phi\right\} $. And it is well-known
the eigen functions and eigen values of such a system are given as,

\begin{align*}
\psi_{n,l,m} & =R_{n,l}\left(r\right)Y_{l,m}\left(\theta,\phi\right),\\
R_{n,l}\left(r\right) & =\mathcal{N}\left(\frac{2r}{na_{0}}\right)^{l}e^{-\frac{r}{na_{0}}}F\left(-n+l+1,2l+2,\frac{2r}{na_{0}}\right),\\
\mathcal{N} & =\frac{2}{a_{0}^{3/2}n^{2}\left(2l+1\right)!}\sqrt{\frac{\left(n+l\right)!}{\left(n-l-1\right)!}},\ E_{n}=-\frac{m_{e}e^{4}}{2n^{2}},\ a_{0}=\frac{1}{m_{e}e^{2}},\tag{B-4}
\end{align*}
where $Y_{l,m}\left(\theta,\phi\right)$ is the 2d spherical harmonic
function and $F\left(a,b,r\right)$ is the hypergeometric function.
Keeping these in hand, let us derive the OTOC $C_{T}^{X},c_{n}^{X}$
and $C_{T}^{Z},c_{n}^{Z}$ by using

\begin{align*}
C_{T}^{X}\left(t\right)=\frac{1}{\mathcal{Z}}\sum_{n}e^{-\beta E_{n}}c_{n}^{X}\left(t\right), & c_{n}^{X}\left(t\right)=-\left\langle n\left|\left[x\left(t\right),p_{x}\right]^{2}\right|n\right\rangle ,\\
C_{T}^{Z}\left(t\right)=\frac{1}{\mathcal{Z}}\sum_{n}e^{-\beta E_{n}}c_{n}^{Z}\left(t\right), & c_{n}^{Z}\left(t\right)=-\left\langle n\left|\left[z\left(t\right),p_{z}\right]^{2}\right|n\right\rangle ,\tag{B-5}
\end{align*}
where $x,z$ refers to the cartesian coordinates given by

\[
x=r\sin\theta\cos\phi,\ z=r\cos\theta.\tag{B-6}
\]
To compute the OTOC, we need the matrix element of $x,z$, hence we
have

\begin{align*}
x_{n^{\prime},l^{\prime},m^{\prime};n,l,m} & =\left\langle n^{\prime},l^{\prime},m^{\prime}\left|r\sin\theta\cos\phi\right|n,l,m\right\rangle \\
 & =\int_{0}^{\infty}drr^{3}R_{n^{\prime},l^{\prime}}^{*}R_{n,l}\int_{0}^{\pi}d\theta\int_{0}^{2\pi}d\phi\sin^{2}\theta\cos\phi Y_{l^{\prime},m^{\prime}}^{*}Y_{l,m}\\
 & \equiv r_{n^{\prime},l^{\prime};n,l}Y_{l^{\prime},m^{\prime};l,m}^{X},\tag{B-7}
\end{align*}
where

\begin{align*}
r_{n^{\prime},l^{\prime};n,l} & =\int_{0}^{\infty}drr^{3}R_{n^{\prime},l^{\prime}}^{*}R_{n,l},\\
Y_{l^{\prime},m^{\prime};l,m}^{X} & =\int_{0}^{\pi}d\theta\int_{0}^{2\pi}d\phi\sin^{2}\theta\cos\phi Y_{l^{\prime},m^{\prime}}^{*}Y_{l,m},\tag{B-8}
\end{align*}
and

\begin{align*}
z_{n^{\prime},l^{\prime},m^{\prime};n,l,m} & =\left\langle n^{\prime},l^{\prime},m^{\prime}\left|r\cos\theta\right|n,l,m\right\rangle \\
 & =\int_{0}^{\infty}drr^{3}R_{n^{\prime},l^{\prime}}^{*}R_{n,l}\int_{0}^{\pi}d\theta\int_{0}^{2\pi}d\phi\sin\theta\cos\theta Y_{l^{\prime},m^{\prime}}^{*}Y_{l,m}\\
 & \equiv r_{n,l;n^{\prime},l^{\prime}}Y_{l,m;l^{\prime},m^{\prime}}^{Z},\tag{B-9}
\end{align*}
where 

\[
Y_{l^{\prime},m^{\prime};l,m}^{Z}=\int_{0}^{\pi}d\theta\int_{0}^{2\pi}d\phi\sin\theta\cos\theta Y_{l^{\prime},m^{\prime}}^{*}Y_{l,m}.\tag{B-10}
\]
Then use the identities for the spherical harmonic function, 
\begin{figure}[h]
\begin{centering}
\includegraphics[scale=0.37]{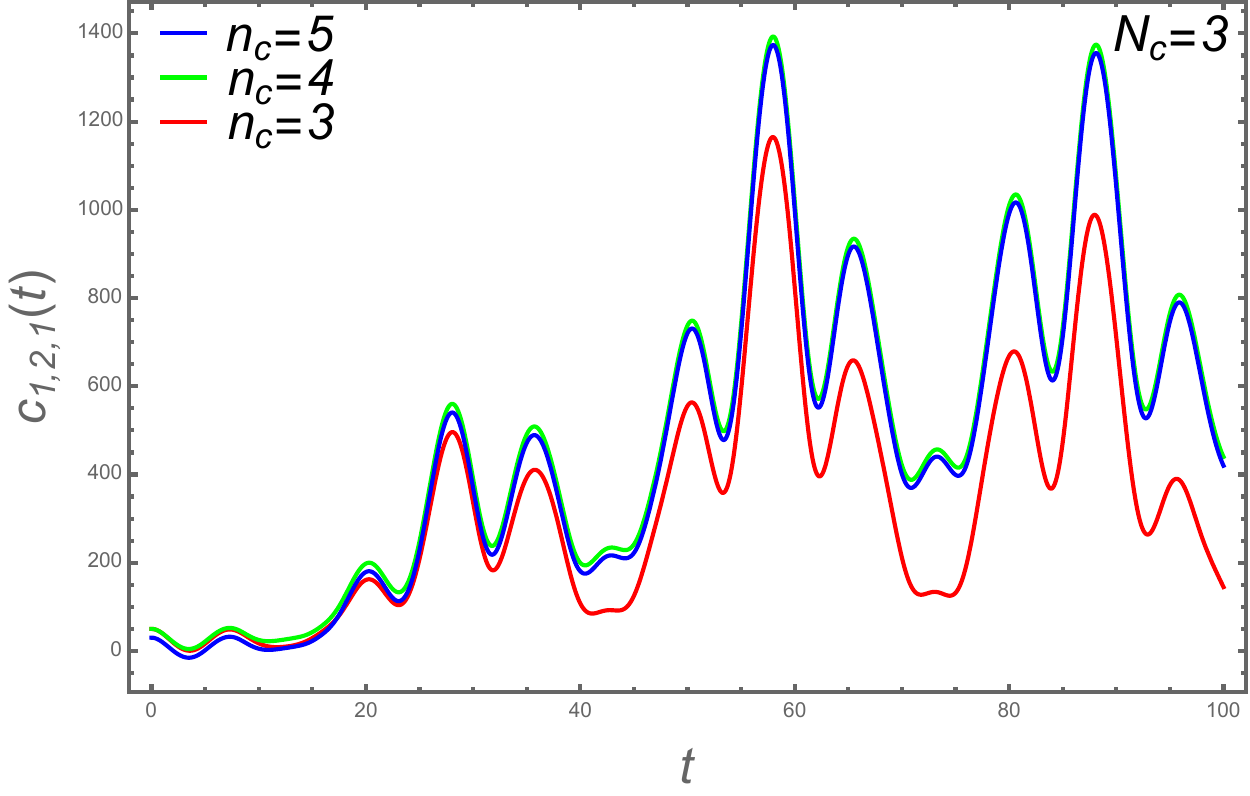}\includegraphics[scale=0.37]{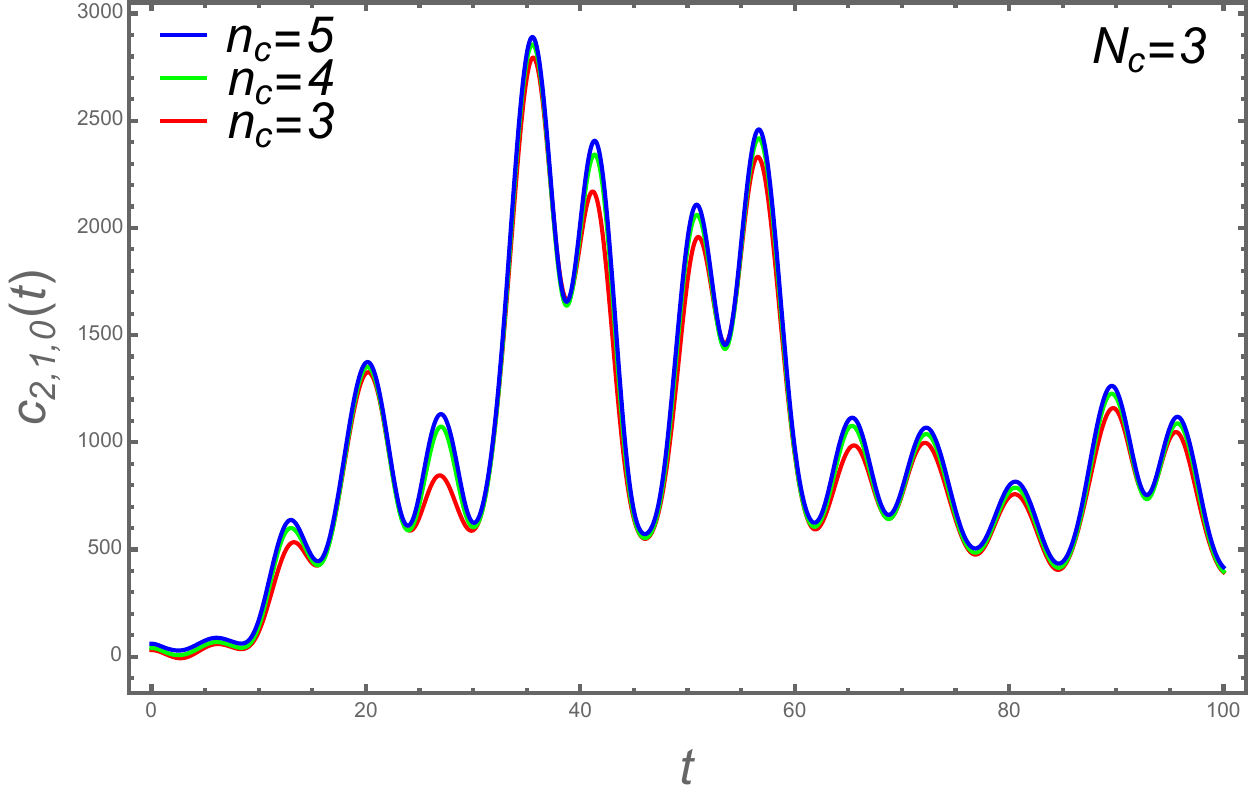}
\par\end{centering}
\begin{centering}
\includegraphics[scale=0.37]{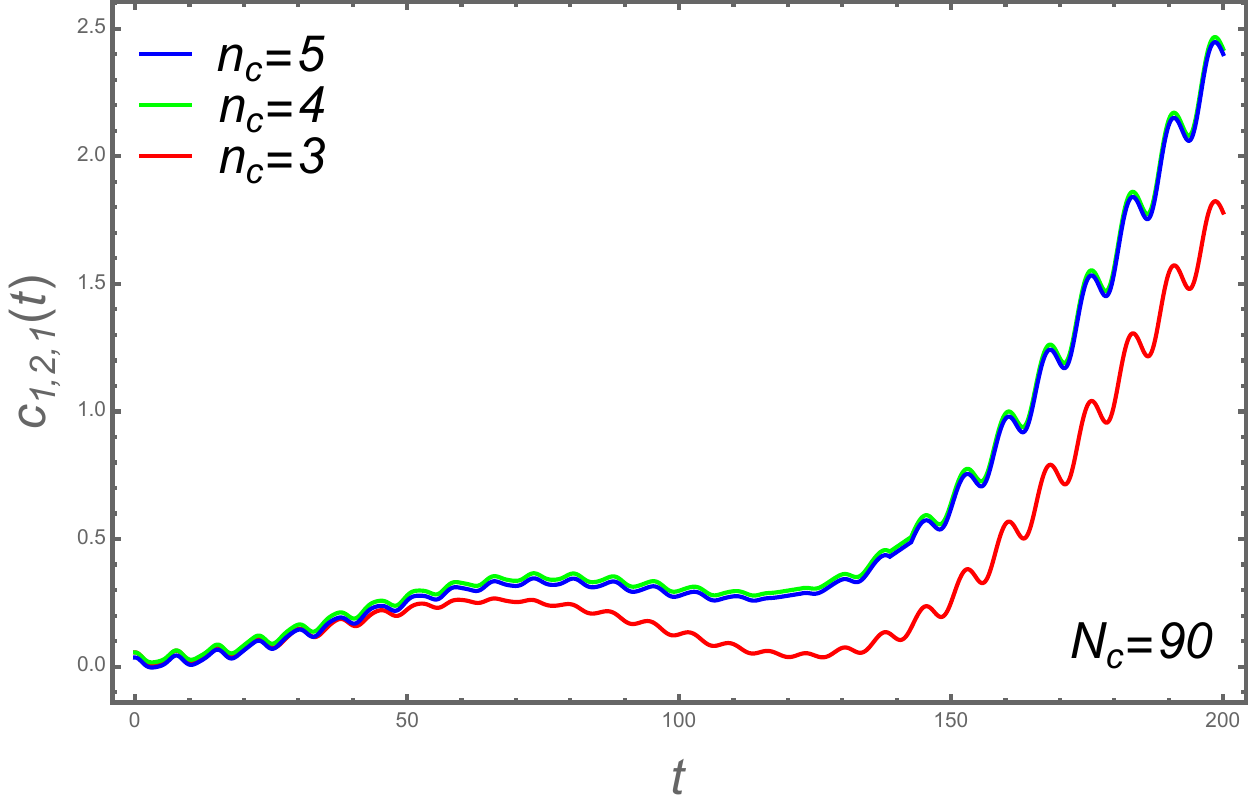}\includegraphics[scale=0.37]{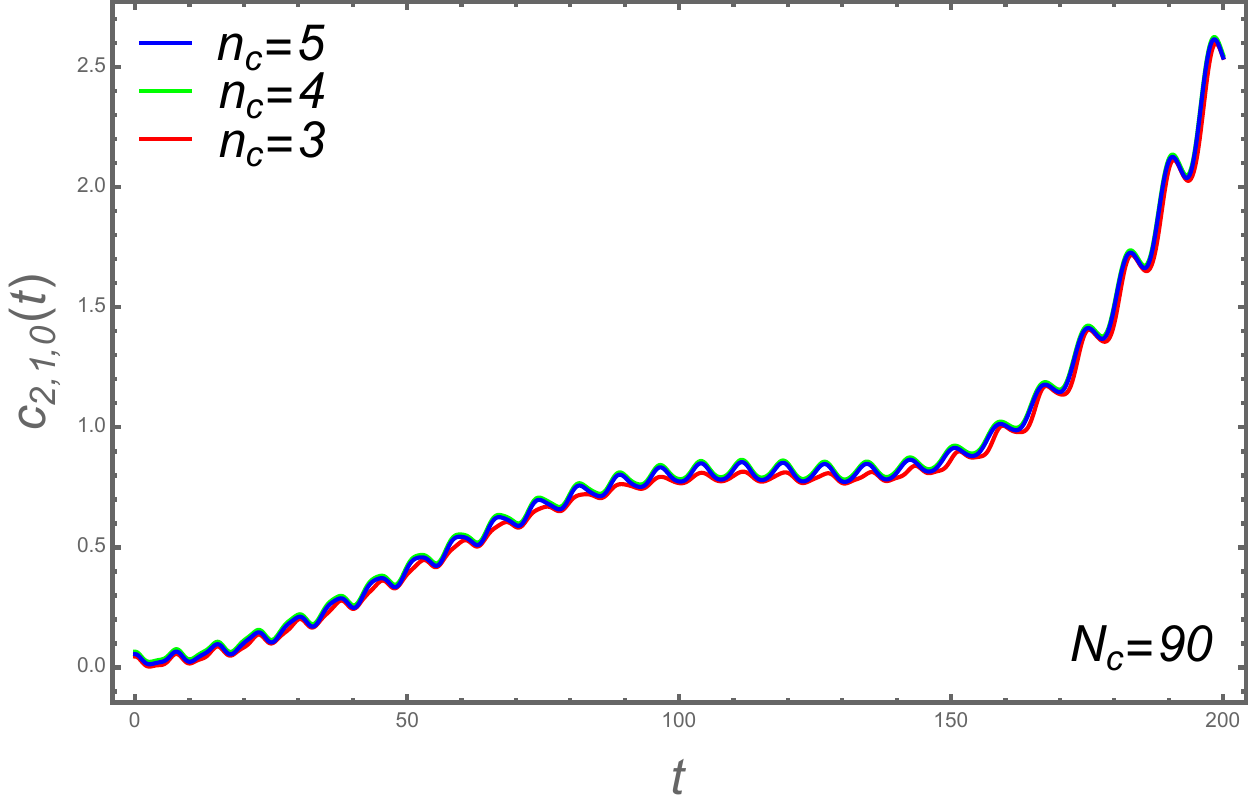}
\par\end{centering}
\caption{\label{fig:C1} The microcanonical OTOC $c_{1,2,1},c_{2,1,0}$ of
holographic Skyrmion with $N_{f}=2$ and $n_{c}=3,4,5$.}
\end{figure}

\begin{align*}
\sin\theta\cos\phi Y_{l,m}= & \frac{1}{2}\bigg[-\sqrt{\frac{\left(l+m+1\right)\left(l+m+2\right)}{\left(2l+1\right)\left(2l+3\right)}}Y_{l+1,m+1}+\sqrt{\frac{\left(l-m\right)\left(l-m-1\right)}{\left(2l-1\right)\left(2l+1\right)}}Y_{l-1,m+1}\\
 & +\sqrt{\frac{\left(l-m+1\right)\left(l-m+2\right)}{\left(2l+1\right)\left(2l+3\right)}}Y_{l+1,m-1}-\sqrt{\frac{\left(l+m\right)\left(l+m-1\right)}{\left(2l-1\right)\left(2l+1\right)}}Y_{l-1,m-1}\bigg],\\
\cos\theta Y_{l,m}= & \sqrt{\frac{\left(l+1\right)^{2}-m^{2}}{\left(2l+1\right)\left(2l+3\right)}}Y_{l+1,m}+\sqrt{\frac{l^{2}-m^{2}}{\left(2l-1\right)\left(2l+1\right)}}Y_{l-1,m},\tag{B-11}
\end{align*}
we can obtain 

\begin{align*}
Y_{l^{\prime},m^{\prime};l,m}^{X}= & \frac{1}{2}\bigg[-\sqrt{\frac{\left(l+m+1\right)\left(l+m+2\right)}{\left(2l+1\right)\left(2l+3\right)}}\delta_{l^{\prime},l+1}\delta_{m^{\prime},m+1}+\sqrt{\frac{\left(l-m\right)\left(l-m-1\right)}{\left(2l-1\right)\left(2l+1\right)}}\delta_{l^{\prime},l-1}\delta_{m^{\prime},m+1}\\
 & +\sqrt{\frac{\left(l-m+1\right)\left(l-m+2\right)}{\left(2l+1\right)\left(2l+3\right)}}\delta_{l^{\prime},l+1}\delta_{m^{\prime},m-1}-\sqrt{\frac{\left(l+m\right)\left(l+m-1\right)}{\left(2l-1\right)\left(2l+1\right)}}\delta_{l^{\prime},l-1}\delta_{m^{\prime},m-1}\bigg],\\
Y_{l^{\prime},m^{\prime};l,m}^{Z}= & \sqrt{\frac{\left(l+1\right)^{2}-m^{2}}{\left(2l+1\right)\left(2l+3\right)}}\delta_{l^{\prime},l+1}\delta_{m^{\prime},m}+\sqrt{\frac{l^{2}-m^{2}}{\left(2l-1\right)\left(2l+1\right)}}\delta_{l^{\prime},l-1}\delta_{m^{\prime},m}.\tag{B-12}
\end{align*}
Since the microcanonical OTOC depends on the matrix elements $Y_{l^{\prime},m^{\prime};l,m}^{X},Y_{l^{\prime},m^{\prime};l,m}^{Z}$,
it means $c_{n}^{X}$ and $c_{n}^{Z}$ are definitely discrepant.
However the thermal OTOC $C_{T}^{X}$ and $C_{T}^{Z}$ should be equivalent
because the thermal OTOC is the average of statistics. To confirm
these comments, we plot out the numerical evaluation of $c_{n}^{X,Z}$
and $C_{T}^{X,Z}$ as it is illustrated in Figure \ref{fig:B}. So
we can see, while $c_{n}^{X}$ could be different from $c_{n}^{Z}$
as it should be, $C_{T}^{X}$ exactly covers $C_{T}^{Z}$ as it is
expected. Beside, our numerical calculation also reveals there is
a time scale that the quantum OTOC $C_{T}$ begin to saturate, however
it does not display the exponential growth either, hence it may also
support that the exponential growth is only the feature of the classical
OTOC. 
\begin{figure}[b]
\begin{centering}
\includegraphics[scale=0.37]{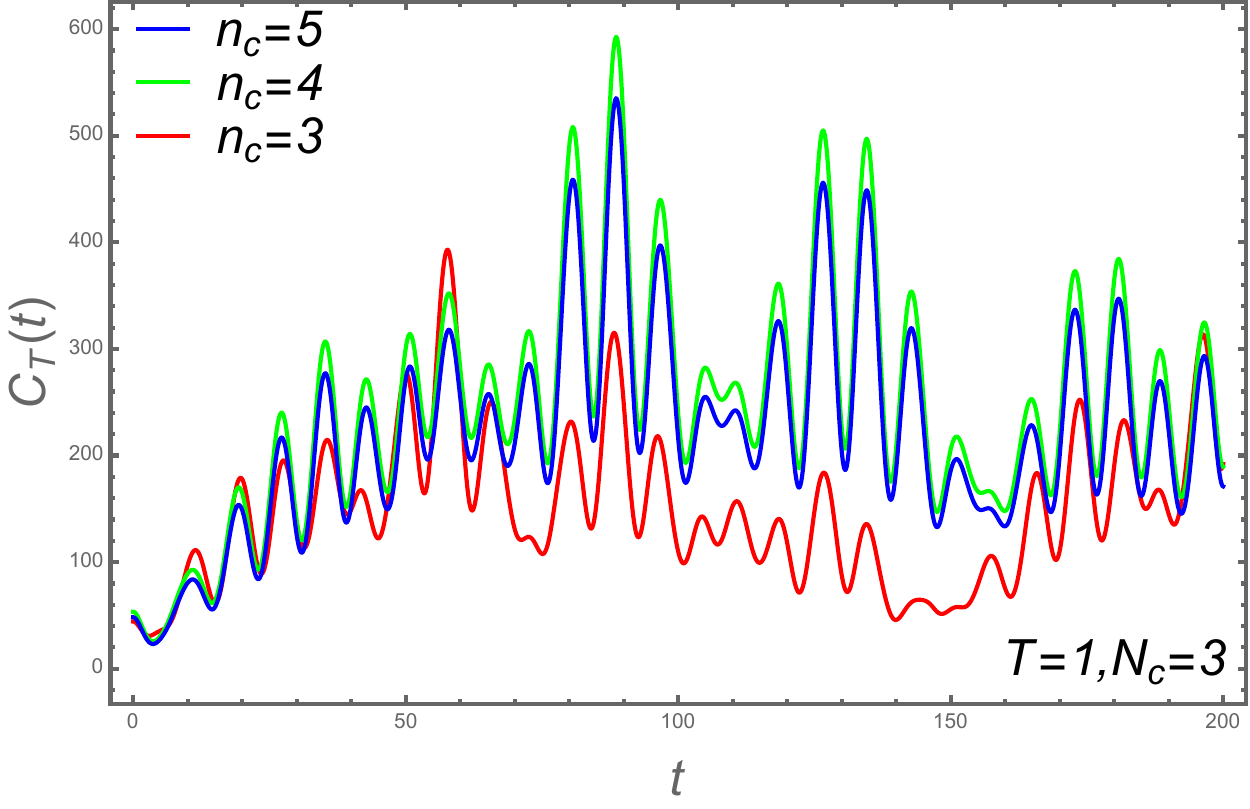}\includegraphics[scale=0.37]{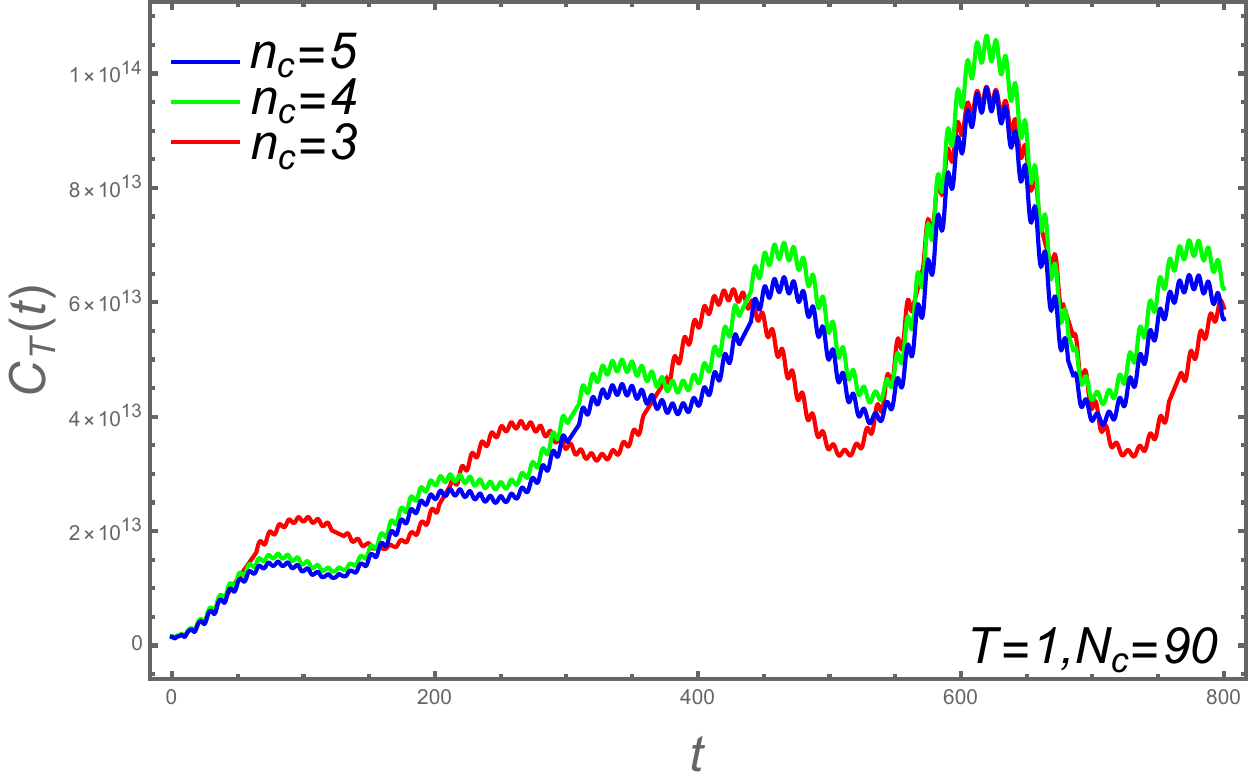}
\par\end{centering}
\caption{\label{fig:C2} The thermal OTOC of holographic Skyrmion at temperature
$T=1$ with $N_{f}=2,N_{c}=3,90$ and $n_{c}=3,4,5$.}
\end{figure}

\section{Appendix C: The truncation error}

As we have displayed, in order to evaluate the OTOCs of the holographic
Skyrmion in the main text, we need to sum infinite terms as it is
shown in Section 2. However, the actual numerical calculation can
not deal with infinite terms in general, hence we have to truncate
the summation at $n_{Z},n_{\rho},l_{n}=n_{c}$. Although there is
no need to include the whole excited states in the OTOC for the realistic
situation, in this appendix let us study the $n_{c}$-dependence for
the completeness in theory. Here, we focus on the two-flavor case
with $N_{c}=3,90$, then consider the OTOC with $n_{c}=3,4,5$. Due
to the degeneracy we have evaluated, it will include respectively
480, 1375, 3276 states contributing to the QTOC. The corresponding
numerical results for the microcanonical OTOCs $c_{1,2,1},c_{2,1,0}$
and thermal OTOCs with various $n_{c}$ are illustrated in Figure
\ref{fig:C1} and \ref{fig:C2}. Accordingly, we can see the OTOC
basically converges when $n_{c}$ increases which implies in our holographic
system the trend of the microcanonical OTOC is not very sensitive
to the truncation. 
\begin{figure}[t]
\begin{centering}
\includegraphics[scale=0.37]{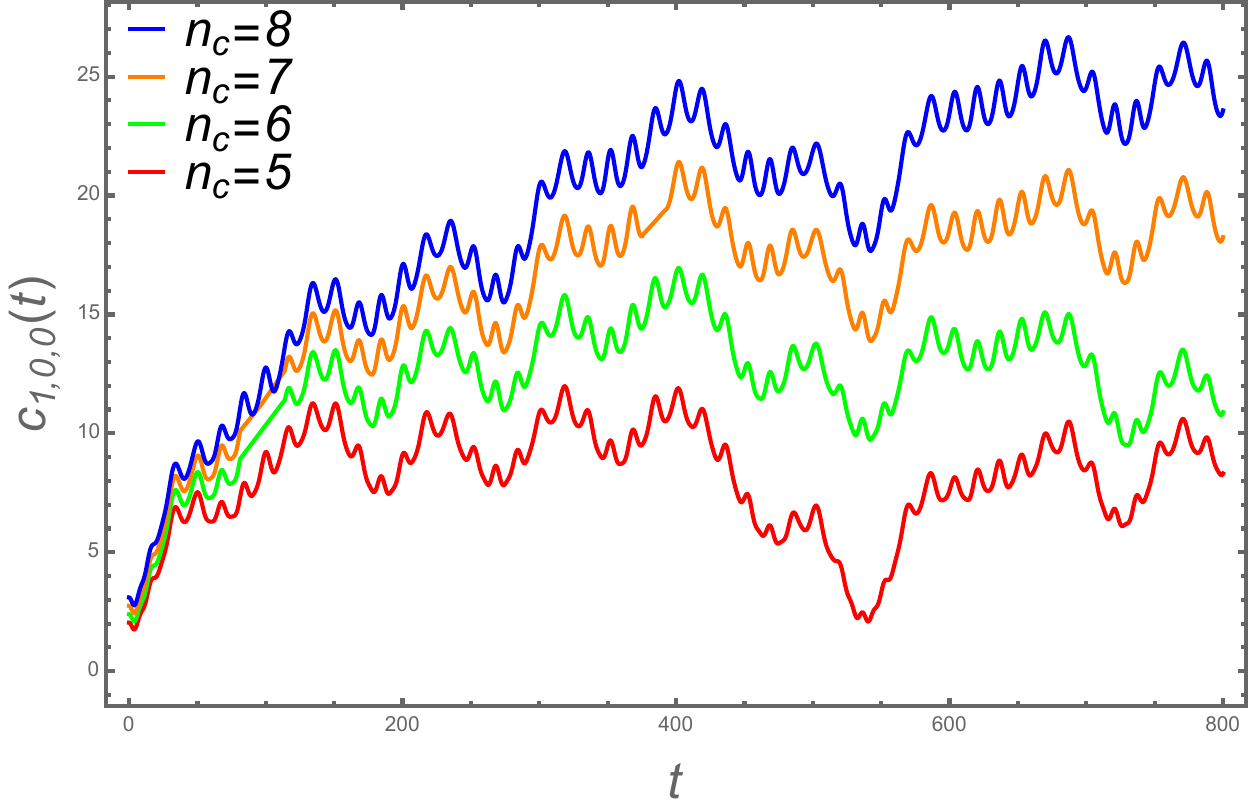}\includegraphics[scale=0.37]{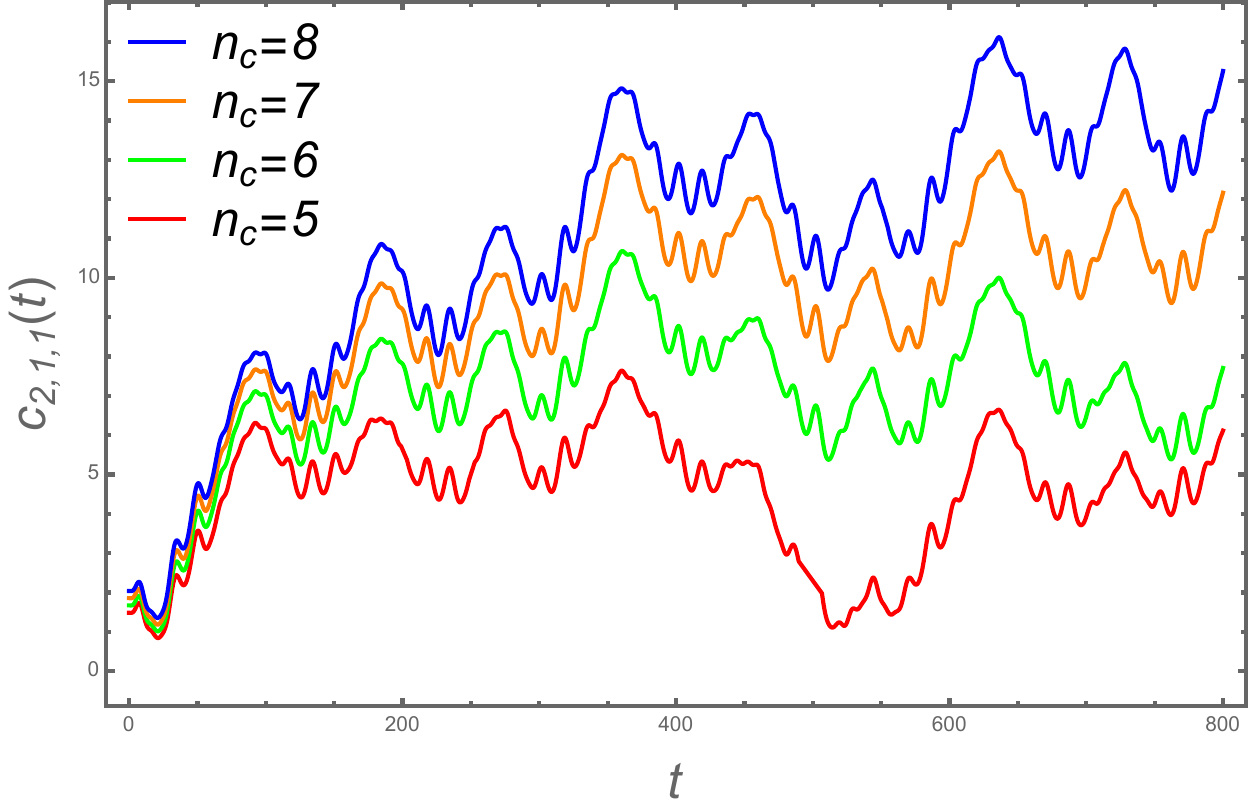}
\par\end{centering}
\begin{centering}
\includegraphics[scale=0.37]{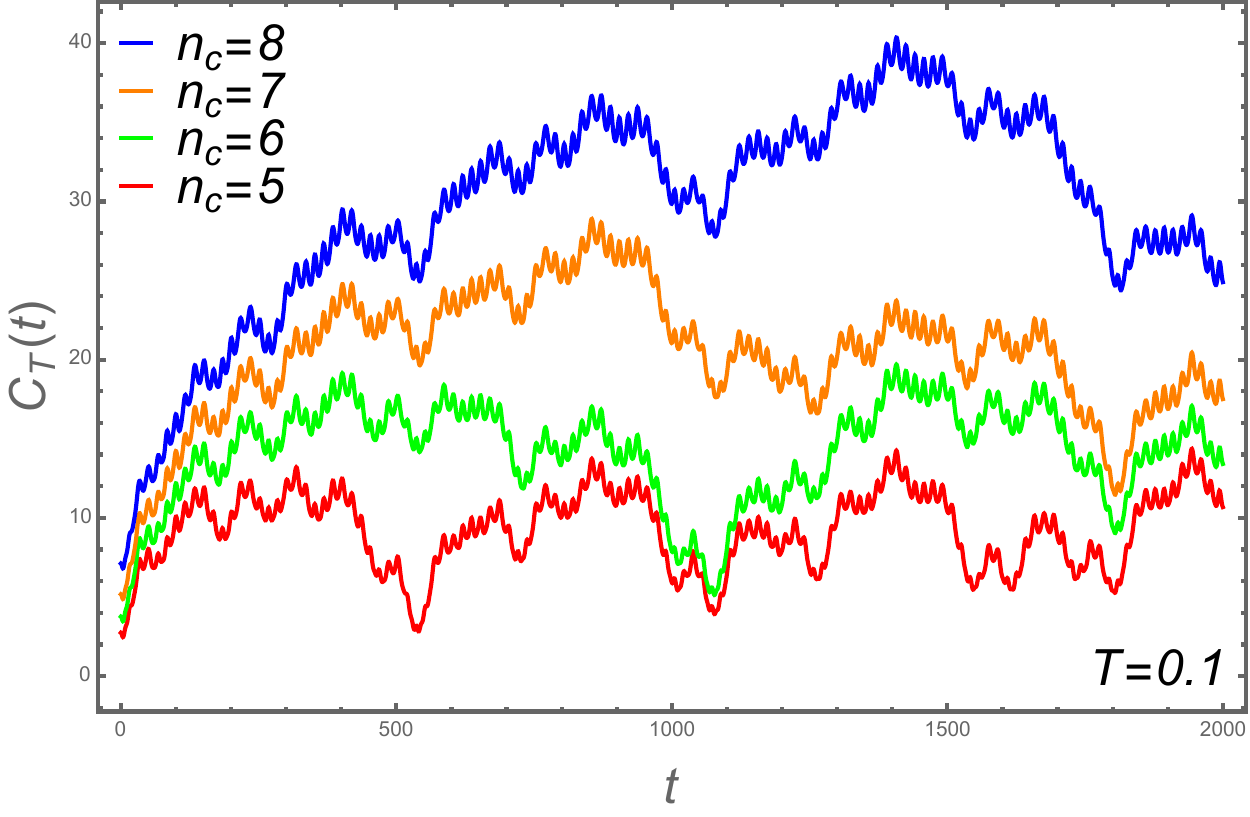}\includegraphics[scale=0.37]{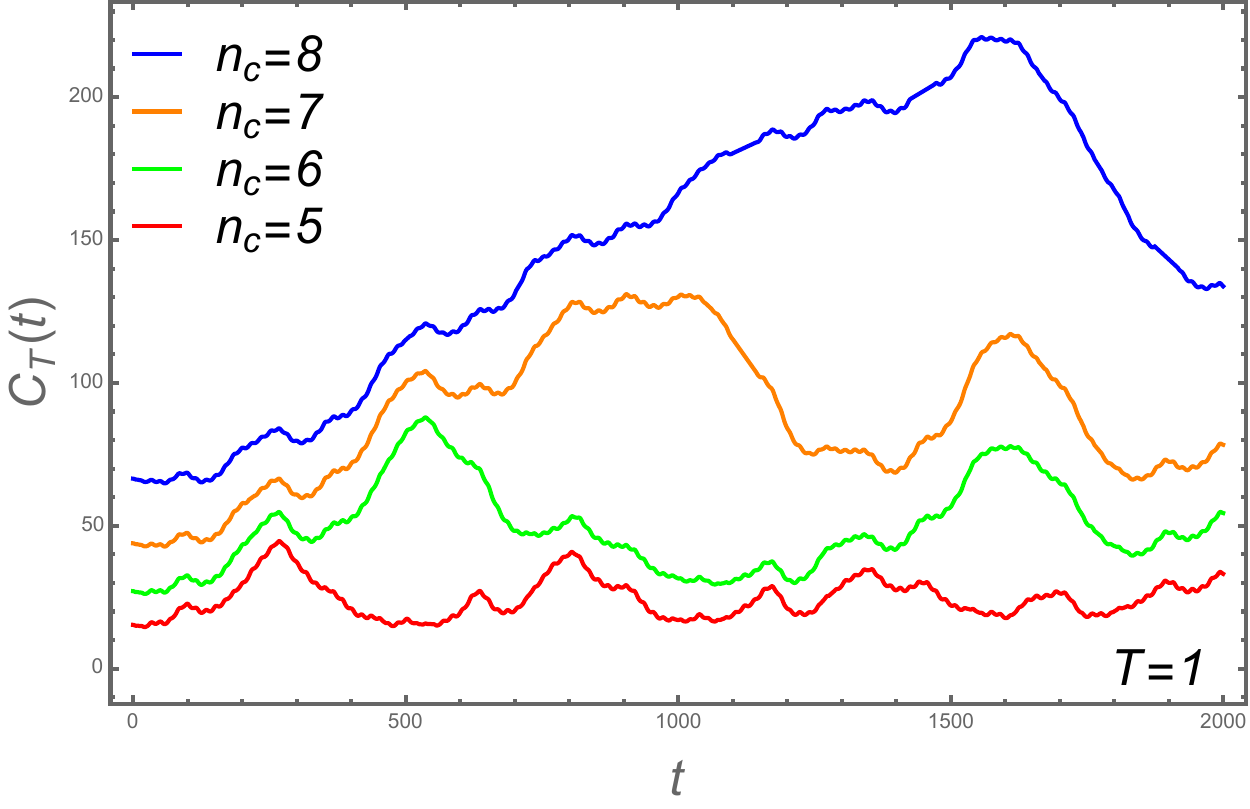}
\par\end{centering}
\caption{\label{fig:C3} The OTOC in quantum mechanics with 3d Coulomb potential
in the unit of $m_{e}=e=1$. \textbf{Upper: }The microcanonical OTOC.
\textbf{Lower: }The thermal OTOC.}
\end{figure}

However, this conclusion may not hold nicely in general. As another
example for the OTOC with the central force field, we have studied
the OTOC with the 3d Coulomb potential in Appendix B, here we plot
numerically out the truncation dependence of the OTOC with the 3d
Coulomb potential in Figure \ref{fig:C3}. The microcanonical OTOC
is denoted as $c_{n,l,m}\left(t\right)$ while the thermal OTOC is
denoted as $C_{T}\left(t\right)$. Therefore, we can see while the
trend of the OTOCs basically remains when the truncation $n_{c}$
increase, OTOCs does not converge very nicely. Nevertheless, it is
possible to study the features of the OTOC with 3d Coulomb potential
if the truncation is sufficiently large.

\end{document}